%% file: paper.tex
\pdfoutput=1
\documentclass[preprint2,numberedappendix,iop]{emulateapj-rtx4}

\usepackage{graphicx,graphics,amsmath}
\usepackage{natbib}
\usepackage{bm,url}
\usepackage{times}
\usepackage{amssymb}
\usepackage{color}
\usepackage{float}
\usepackage{arydshln}
\graphicspath{{./fig/}{./png/}}
\include{newcommands}

\usepackage{color}

\shorttitle{Solar Cycle Variability in a Babcock-Leighton Dynamo Model}
\shortauthors{Karak \& Miesch}

\clearpage
\begin{document}

\title{Solar Cycle Variability Induced by Tilt Angle Scatter in a Babcock--Leighton Solar Dynamo Model}
\author{Bidya Binay Karak and Mark Miesch}
\affil{High Altitude Observatory, National Center for Atmospheric Research, 3080 Center Green Dr., Boulder, CO 80301, USA}
\begin{abstract}
We present results from a three-dimensional Babcock--Leighton dynamo model 
that is sustained by the explicit emergence and dispersal of bipolar magnetic regions (BMRs).  
On average, each BMR has a systematic tilt given by Joy's law. 
Randomness and nonlinearity in the BMR emergence of our model produce variable magnetic 
cycles.
However, when we allow for a random scatter in the tilt angle to mimic the observed departures from Joy's law,
we find more variability in the magnetic cycles. 
We find that the observed standard deviation in Joy's law of $\sigma_\delta = 15^\circ$ 
produces a variability comparable to observed solar cycle variability of $\sim $ 32\%, 
as quantified by the sunspot number maxima between 1755--2008.  We also find that tilt angle scatter can 
promote grand minima and grand maxima.  The time spent in grand minima for $\sigma_\delta = 15^\circ$ is somewhat less than that inferred for the Sun from cosmogenic isotopes (about 9\% compared to 17\%).  However, when we double the tilt scatter to $\sigma_\delta = 30^\circ$, the simulation statistics are comparable to the Sun ($\sim $18\% of the time in grand minima 
and $\sim 10$\% in grand maxima).  
Though the Babcock--Leighton mechanism is the only source of poloidal field, 
we find that our simulations always maintain magnetic cycles 
even at large fluctuations in the tilt angle.
We also demonstrate that tilt quenching is a viable and efficient mechanism for dynamo saturation; a suppression of the tilt by only 1-2$^\circ$ is sufficient to limit the dynamo growth.  Thus, any potential observational signatures of tilt quenching in the Sun may be subtle.
\end{abstract}

  \email{bkarak@ucar.edu}
  \maketitle

\section{Introduction}
\label{sec:int}
The $11$-year solar cycle is a manifestation of the oscillatory magnetic field of the Sun.
The solar cycle, however, is not regular. The strength,
as well as the period, have an irregular variation. The extreme
example of such variation is the Maunder minimum in the $17$th
century when sunspots largely disappeared for about $70$~years.  Indirect
studies suggest that there were many such events in the past
\citep{Uso13}.

There is no doubt that a dynamo mechanism, operating in the 
solar convection zone (SCZ),
is responsible for producing the solar magnetic cycle. Thus the
natural way of studying the solar dynamo is by solving the basic
magnetohydrodynamic (MHD) equations in a rotating spherical shell,
encompassing the SCZ. 
However, though substantial progress has been made in recent years in studying
fundamental dynamo mechanisms \citep[e.g.,][]{C14,ABMT15,FM15,HRY16,Kap16,KB16}, MHD 
simulations still cannot capture all processes relevant to the solar dynamo
and the solar cycle \citep{FF14,Kar15}. One
reason could be that
these simulations do not produce sufficient flux emergence in the
form of tilted bipolar magnetic regions (BMRs) that we see in the solar observations
\citep[e.g.,][]{WS89}. These tilted BMRs, when they decay
and disperse on the solar surface, produce a large-scale poloidal field, 
as proposed by \citet{Ba61} and \citet{Le64}.
Recent high-quality BMR (area, tilt, separation, etc) 
and polar field data (measured both directly via polarization and 
indirectly through different proxies, including polar faculae and 
active networks), suggest that this process is sufficient to maintain
the observed polar flux in the Sun \citep{Das10,KO11,Muno13,Priy14}.

In the Babcock-Leighton (BL) paradigm, the poloidal flux 
produced by the decay of tilted BMRs gets transported downward,
to the bulk of the SCZ, by meridional circulation and convection.
There the differential rotation stretches this poloidal field to
produce a toroidal component---the $\Omega$ effect. 
This toroidal flux then produces BMRs on the surface 
consistent with the Hale polarity rule \citep{Hale19,SK12}, although 
there are some difficulties in constraining how and where BMRs
are formed.  By comparing the observed magnetic flux on the solar
surface with the flux generated by the differential rotation,
\citet{CS15} have 
argued that the $\Omega$ effect 
can account for the toroidal flux that ultimately emerges 
as BMRs.  This suggests that
the solar dynamo is of
the $\alpha_{\it BL} \Omega$ type, where the $\alpha_{\it BL}$ is the
symbol for the BL process. Following this basic dynamo
loop, and using the turbulent diffusivity and meridional flow for
the flux transport, many authors have developed 2D as well as 3D
\bl\ dynamo models \citep[see reviews by][]{Cha10,Kar14a}. Most of
these models have been able to reproduce the basic features of the
solar cycle.

Possible causes of solar cycle variability in the \bl\ dynamo 
framework include variations in
(i) convective transport, (ii) the meridional circulation, (iii)
the differential rotation, and (iv) the \bl\ process.
Note that Lorentz forces play a role in all of these mechanisms 
so they are not listed as a separate item.

Flux transport by convective flows (i) 
definitely has stochastic elements and nonlinear feedbacks due to 
the dynamo-generated magnetic field and has been studied by some 
authors \citep[e.g.,][]{KPR94,Kar14b}.  
However, this is a challenging
problem that will require a unified understanding of small and 
large-scale dynamo action to fully address.
The influence of the meridional circulation in particular (ii) has been
investigated by a number of authors and has been shown to give rise
to cycle variability, including grand minima and grand maxima
\citep[e.g.,][]{CD00,LP09,KC11,KC13,UH14}. Weak variations in 
the differential rotation (iii) are known to exist, namely 
torsional oscillations.  However, the observed correlation
between the polar flux at cycle minimum and the sunspot number
of the following cycle suggests that the $\Omega$-effect may
be largely linear and therefore not a major source of cycle
variability \citep{JCC07,WS09,Muno13}.

Here we focus on mechanism (iv), namely the variability 
induced by the BL process. The poloidal field generated in
this process largely depends on the amount of flux in BMRs,
the frequency of BMR eruptions, and the tilt angles of BMRs.
All of these quantities have temporal variations.  Since,
on average, there are only about two new BMRs per day on the
solar surface, the fluctuations in any of these quantities
can lead to a considerable variation in the poloidal field and
consequently the cycle strength. 

In this work, we investigate mechanism (iv) in an innovative way, using our 3D 
STABLE (Surface flux Transport And Babcock--LEighton) solar dynamo
model \citep[][hereafter MD14, and MT16, respectively]{MD14,MT16}.
We focus in particular on the influence of the observed tilt angle 
distribution by superposing a random scatter on the Joy's law 
prescription that we used in previous work. We have also introduced 
tilt angle quenching into STABLE as a mechanism for dynamo saturation.

In addition to implementing tilt angle scatter, we have also 
modified the flux distribution of BMRs.  Previously, the flux
of each BMR was directly linked to the low-latitude toroidal flux at the base of the CZ.
In this study, we improve the 
realism of the model by choosing a BMR flux distribution based on solar 
observations.  Furthermore, we consider two alternative ways to regulate the 
photospheric flux budget.  The first is to increase the amount of magnetic
flux in each BMR by shifting the flux distribution toward larger values
when the toroidal flux near the base of the CZ is large.
The second is to keep the flux distribution the same and vary the rate of BMR emergence
in response to the toroidal flux near the base of the CZ.  In this latter approach,
the range of emergence rates we use is consistent with solar observations.

In our study, we first ask
several questions, namely, whether the solar dynamo can be
maintained through the observed properties of BMRs without
any other source of the poloidal field, whether the quenching
in the tilt angle is sufficient to saturate the dynamo, and how
robust this model is with different algorithms of BMR deposition
frequency and with different values of diffusivity.  Then, we explore
the variation of the magnetic cycle due to the 
observed variation in the BMR tilt angles.

Random scatter in BMR tilt angles has been proposed by many authors
as a possible mechanism to explain the irregularity of the solar
cycle and it has been studied previously within the context of
2D BL dynamo models \citep[e.g.,][]{CD00,JCC07,CK09,OK13}, surface 
flux transport (SFT) models
\citep{JCS14,HU16} and in a coupled $2$D$\times2$D BL/SFT dynamo
model \citep{LC16}. 
However, to our knowledge, our model is the first 3D solar dynamo model to explicitly investigate the implications of tilt angle scatter with regard to solar cycle variability.

After analyzing the features of magnetic cycles
obtained from this model, we explore whether the variation
in the tilt angle can also lead to the extreme cycle modulation such as grand minima and maxima.
Finally, we explore the robustness of our model,
and in particular, whether it continues to 
produce 
magnetic cycles 
when the tilt angle scatter becomes very large.

\section{Model}
\label{sec:mod}
In our model, we solve the induction equation,
\begin{equation}
\frac{\partial {\bm B}}{\partial t} =  {\bf \nabla} \times \{ ({\bm V + {\bm \gamma}}) \times {\bm B} - \eta_{\rm t} \nabla \times {\bm B}\},
\label{eqind}
\end{equation}
in three dimensions ($r$, $\theta$, $\phi$) 
for the whole SCZ 
with $0.69R \le r \le R$ (= radius of the Sun), $0 \le \theta$ (colatitude) $\le \pi$, and $0 \le \phi \le 2\pi$.
In the simulations reported here, our model is kinematic and the velocity 
field $\bm V$ is composed of axisymmetric 
meridional circulation ($v_r$ and $v_\theta$) and differential rotation ($v_\phi/r \sin\theta$), such that
\begin{equation}
{\bm V} = v_r(r,\theta) \rhat + v_\theta(r,\theta) \thetahat + r \sin \theta~\Omega(r,\theta) \phihat.
\label{eqv}
\end{equation}

\begin{figure}
\centering
\includegraphics[width=1.0\columnwidth]{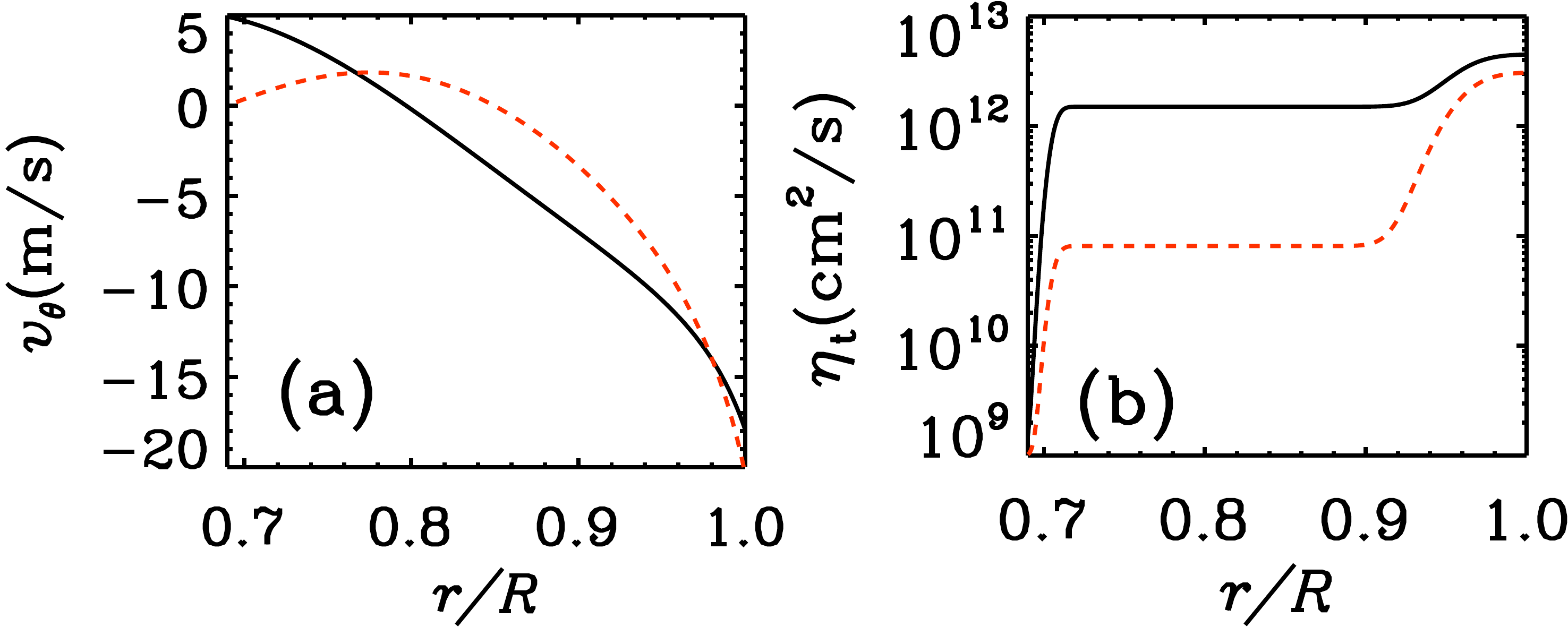}
\caption{The radial variations of (a) the latitudinal component of the velocity $v_\theta$
at mid-latitude and (b) the turbulent diffusivity $\etat$
used in the advection-dominated model (dashed/red line)
and the diffusion-dominated model (solid line).
}
\label{fig:mc}
\end{figure}

For the meridional circulation, we use the profile given in many previous publications, particularly
in \citet{KC16} (Equation 5) which closely resembles the surface observations. 
Hence, without repeating the mathematical equations of this flow we just make a few comments: 
near the 
surface it is poleward with a maximum speed of $20$~\mps, near the base of the CZ it is equatorward 
with a speed of about $2$~\mps, and it 
smoothly goes to zero at the lower boundary ($0.69R$); see dashed line in \Fig{fig:mc}(a). 

Note that in this study we have considered a single cell circulation.  Recent helioseismic inversions suggest that this may not be accurate but they have not yet converged on a robust determination of what the structure and amplitude may in fact be \citep{jacki15,RA15,ZC16}.  In the absence of this information and to make contact with previous BL dynamo models, we have retained the single-celled profile.  Others have investigated the role of multi-celled circulation profiles in 2D BL/flux transport dynamo models and they have demonstrated that these models are still viable, provided that the circulation near the base of the convection zone is equatorward and that the convective transport of poloidal flux (typically parameterized by a turbulent diffusion and a magnetic pumping) is sufficiently efficient \citep{jouve07,HKC14,beluc15,hazra16}.

For differential rotation, we use an analytic 
function that captures the observed helioseismic data. This profile has been used in many
previous publications, for example, see Equation~(3) of MT16.

The $\bm \gamma$, appearing as an advective term in \Eq{eqind}, is the magnetic pumping. 
In most of the simulations, we include a
downward magnetic pumping motivated by the study of \citet{KC16}.
Thus we write $\bm \gamma = \gamma_r(r) \rhat$, where
\begin{eqnarray}
\gamma_r(r) = - \frac{ \gamma_{\rm CZ} }{2}\left[1 + \mathrm{erf} \left(\frac{r - 0.725R}{0.01R}\right) \right] \nonumber  \\
              - \frac{\gamma_{\rm S}}{2}\left[1 + \mathrm{erf} \left(\frac{r - 0.9R}{0.02R}\right) \right].
\label{eqpump}
\end{eqnarray}
Due to the lack of knowledge of the exact latitudinal variation of $\gamma_r$ 
we take it to be only a function of radius.
As discussed in \citet{KC16}, the pumping is efficient near the surface (mainly caused by the 
topological asymmetry of the convective flow), while the deeper convection is weaker and less stratified  \citep{Spr97}.  The pumping helps to boost the efficiency of the dynamo by suppressing the diffusion of toroidal flux through the surface.
The the amount of pumping used in each simulation varies depending on the value of diffusivity used. 
Hence \vpCZ\ and \vpS\ 
will have different values in different simulations.

In the present model, we do not consider the small-scale convective flow and thus 
to capture its mixing effect we consider an effective 
turbulent diffusivity represented by $\etat$ in \Eq{eqind}. This is actually the sum of the molecular 
and turbulent diffusivities.
We do not have a reliable estimate of $\etat$ in the deep CZ. The mixing length theory and 
other theoretical studies suggest that 
the value of $\etat$ in the mid convection zone is of the order of $10^{12}$~cm$^2$~s$^{-1}$ \citep{Par79,Miesch12,CS16,SCD16}. 
Near the surface, at least, it is 
fairly constrained by observations as well as
by the surface flux transport model \citep[e.g.,][]{KHH95,LCC15} 
and it is about a few times $10^{12}$~cm$^2$~s$^{-1}$.
Hence in our model, we choose the following radial dependent profile for $\etat$:
\begin{eqnarray}
\etat(r) = \etaRZ + \frac{\etab}{2}\left[1 + \mathrm{erf} \left(\frac{r - 0.715R}{0.0125R}\right) \right]+ \nonumber  \\
\frac{\etas}{2}\left[1 + \mathrm{erf} \left(\frac{r - 0.956R}{0.025R}\right) \right],
\label{eqeta}
\end{eqnarray}
where
$\etaRZ = 1.0 \times 10^9$~cm$^2$~s$^{-1}$, and $\etas = 3 \times 10^{12}$~cm$^2$~s$^{-1}$.
We have broadly two sets of simulations. In one set, $\etab = 5 \times 10^{10}$~cm$^2$~s$^{-1}$,
while in the other set, $\etab = 1.5 \times 10^{12}$~cm$^2$~s$^{-1}$; see \Fig{fig:mc}(b). Thus the first
set of simulations will be close to our previous publications (MD14, MT16) 
in terms of the diffusion while the latter set will be in the so-called diffusion-dominated regime 
where diffusive flux transport across the CZ dominates over advection by the meridional circulation 
\citep[e.g.,][]{JCC07,YNM08}.

A major component of our model is the SpotMaker algorithm which deposits BMRs on the surface
based on the toroidal flux near the base of the convection zone. 
In SpotMaker, we do the following steps. First, we
compute the strength of the spot-producing toroidal flux near the base of the CZ
\begin{eqnarray}
 \hat{B}(\theta,\phi,t) = \int_{r_a}^{r_b} h(r) B_\phi (r,\theta,\phi,t) dr,
\label{eqspotprodflc}
\end{eqnarray}
where, $r_a=0.715R$, $r_b=0.73R$, and $h(r)=h_0(r-r_a)(r_b-r)$ with $h_0$ as a normalization factor.
We note that to have a prominent equatorward migration of sunspots, the spot-producing toroidal flux 
is computed above the tachocline where the flow is strongest.
A necessary (but not sufficient) condition to produce a BMR is that $\hat{B}(\theta,\phi,t)$ exceeds a threshold field strength
$B_t(\theta)$.
If this condition is satisfied on multiple grid points, then out of those points randomly one point is chosen.
Unlike previous publications (MD14, MT16) where a fixed value was taken for this threshold field strength,
here we make it latitude dependent such that it increases exponentially towards the higher latitudes.
Hence we choose
\begin{eqnarray}
 B_t(\theta) = B_{t0} \exp\left[\gamma (\theta-\pi/2) \right], \quad
 \mathrm{for}~~ \theta  > \pi/2 \nonumber\\
 = B_{t0} \exp\left[\gamma (\pi/2-\theta) \right], \quad
 \mathrm{for}~~ \theta  \le \pi/2
\end{eqnarray}
where $\gamma =5$ and $B_{t0} = 2~$kG.
The rapid increase of $B_t$ in latitude is chosen to have sufficient spots near the equator 
and no spots beyond about $\pm 30^\circ$ latitudes. 
The advantage of using such latitude dependent $B_t$ is that now we do not have to choose any arbitrary 
masking function to suppress spots above a certain latitude
which was used in many previous works \citep[e.g.,][MD14]{Dik04}.
Another advantage
is that now the upper latitudinal bound for BMR emergence is not fixed and 
it can vary depending on the toroidal field strength in each cycle and even in each hemisphere. 
This is consistent with observations that stronger cycles start producing sunspots at 
slightly higher latitudes \citep{SWS08}.
Other than some tachocline instabilities which might be operating in higher latitudes
to destabilize the spot-producing toroidal field \citep{GD00,PM07,Dik09},
we have to confess that, at the moment, we do not have a clear understanding 
of why BMRs do not
appear above a certain latitude and the arbitrary masking function or the latitude dependent $B_t$
chosen here may be regarded as a semi-empirical model.

\begin{figure}
\centering
\includegraphics[width=0.75\columnwidth]{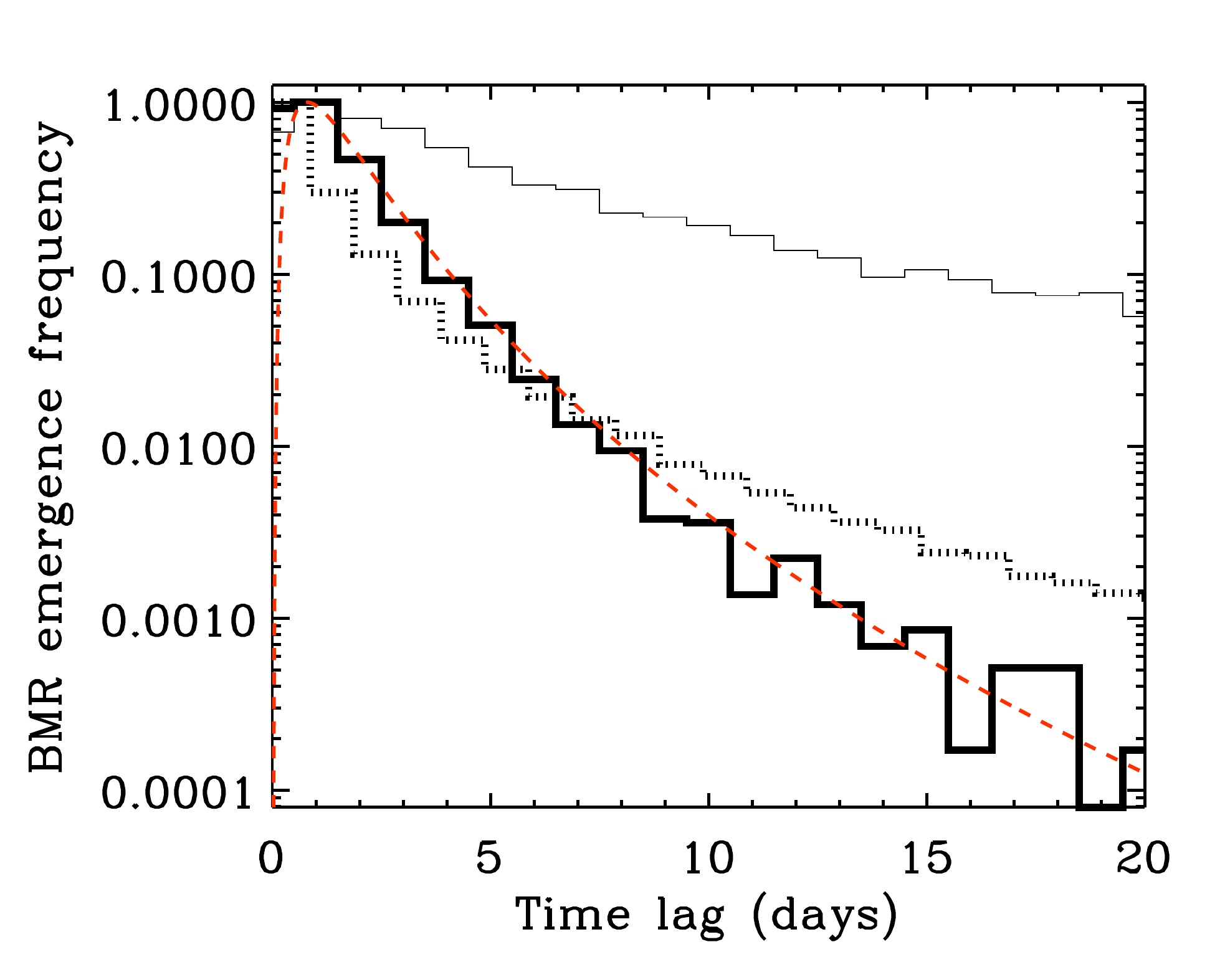}
\caption{Normalized histograms of the time delays between the successive BMR emergences 
obtained from the observed data during $1900$--$2002$. The thick solid line is obtained 
by taking data within a three-year window at each cycle maximum, while the thin solid line
represents the rest of data, i.e., covering the solar minimum periods. The dashed/red line 
is the fitted log-normal distribution with \{$\taup, \taus\} \equiv \{0.8, 1.9$\}~days as given by \Eq{eq:delaypdf}. 
The dotted line is obtained from our model (Run~B9), in which the time delay is related to the magnetic field through \Eq{eqtau} in \Sec{sec:BdepTD}.
}
\label{fig:tdelay}
\end{figure}

When SpotMaker produces a BMR, we do not reduce the flux locally at the progenitor location
although we do place opposing flux near the surface by virtue of the 3D structure of the BMRs;
see Section~2.3 of MT16 for details on this issue. 
Therefore, at every time step of our numerical integration, 
if the BMR emergence is determined only by the criterion
$\hat{B}(\theta,\phi) > B_t(\theta)$, then we may have BMRs emerging at every time step and the total number of BMRs
will largely be determined by the integration time step and the value of $B_{t0}$. 
Thus, to make the emergence rate independent of the numerics and more realistic,
we specify a time delay between two successive BMRs based on solar observations.
The time delay distribution obtained from the observed sunspot data 
(Royal Observatory Greenwich -- USAF/NOAA Sunspot\footnote{Compiled by David Hathaway, http://solarscience.msfc.nasa.gov/greenwch.shtml})
during $1900$--$2002$ is shown by the thick solid line in \Fig{fig:tdelay}.
We approximate this data by a log-normal distribution given by
\begin{eqnarray}
P(\Delta) = \frac{1}{ \sigma_{\rm d} \Delta \sqrt{2\pi} } \exp\left[- \frac{(\ln\Delta - \mu_{\rm d})^2}{2\sigma_{\rm d}^2} \right],
\label{eq:delaypdf}
\end{eqnarray}
where $\sigma_{\rm d}$ and $\mu_{\rm d}$ are specified in terms of the mean $\taus$ and mode $\taup$ 
of the distribution such that $\sigma_{\rm d}^2 = (2/3) \left[ \ln\taus - \ln\taup \right] $ 
and $\mu_{\rm d} = \sigma_{\rm d}^2 + \ln\taup$.
When $\taup = 0.8$~days and $\taus=1.9$~days, the above log-normal distribution 
reasonably fits the observed data as shown by the red/dashed line in \Fig{fig:tdelay}.
We, however, note that the observed time delay shown by the thick solid line in \Fig{fig:tdelay} 
is obtained only from the three years data during each solar maximum,
and not from the full period of $1900$--$2002$ data. Actually, during the solar minimum, we observe
a less frequent BMR and the time delay is much longer; see the thin solid line in \Fig{fig:tdelay}. 
Thus the time delay, in reality, is cycle-phase dependent---it is shortest at the peak of the cycle and longest
at the minimum. However, as a first step, we shall perform a set of simulations by taking fixed values of 
$\taup = 0.8$~days and $\taus=1.9$~days,
obtained from the solar maxima data. Later in \Sec{sec:BdepTD}, 
we shall implement a solar cycle dependent time delay
by considering $\taup$ and $\taus$ as the toroidal field dependent.
We note that the time delay in each hemisphere is always computed separately using \Eq{eq:delaypdf} 
so that no hemispheric symmetry is imposed in this process.

Just to summarize the whole idea, SpotMaker produces the first BMR once the condition 
$\hat{B}(\theta,\phi) > B_t(\theta)$ is satisfied.
Then after a time $dt$ since the time of the previous BMR appearance, 
the SpotMaker produces the next BMR only when both conditions, 
$\hat{B}(\theta,\phi) > B_t(\theta)$
and $dt \ge \Delta^{N(S)}$, where $\Delta^{N(S)}$ is the time delay randomly obtained from the long-normal distribution given by \Eq{eq:delaypdf} 
for northern (or southern) hemisphere.
The superscript ``$N(S)$" on $\Delta$ is to emphasize that the time delay between BMRs 
can be different in two hemispheres as the probability is computed separately in two hemispheres.

\begin{figure}
\centering
\includegraphics[width=0.70\columnwidth]{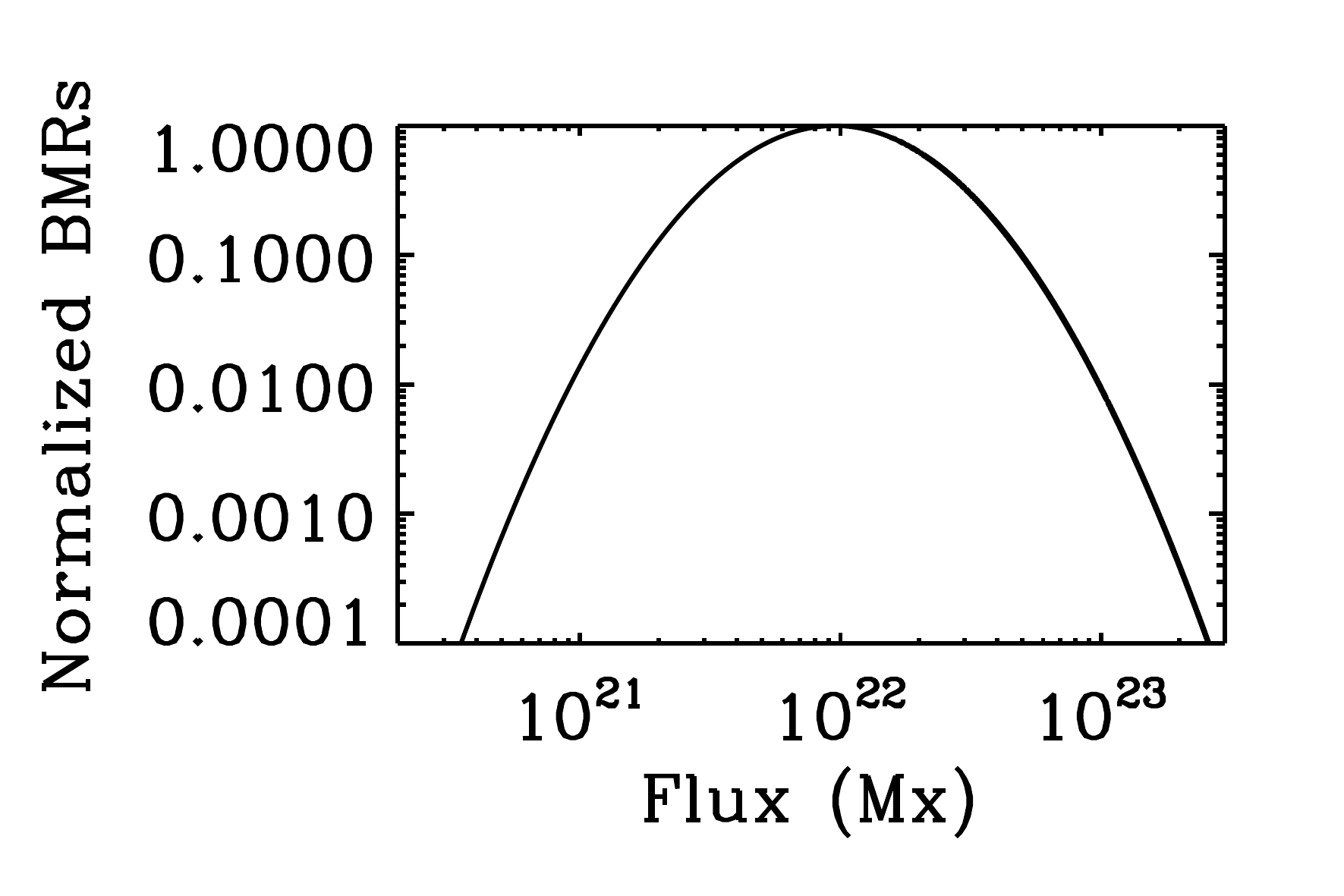}
\caption{Normalized flux distribution of BMRs used in our dynamo model.}
\label{fig:fluxdist}
\end{figure}

Once SpotMaker decides to produce a BMR on the surface, we need to specify its flux, tilt, 
separation and spatial distribution.
In comparison to previous publications \citep[MD14, MT16,][]{HCM17}, here we have some changes in order to
make a close connection with observations. In the previous model, the BMR flux 
was directly related to the toroidal field at the base of the CZ while in this model,
it is obtained from the observed distribution. The observed BMR flux distribution can be approximated using a
log-normal distribution:
\begin{eqnarray}
P(\Phi) = \Phi_0 \frac{1}{ \sigma_\Phi \Phi \sqrt{2\pi} } \exp\left[- \frac{(\ln\Phi - \mu_\Phi)^2}{2\sigma_\Phi^2} \right],
\label{eqflux}
\end{eqnarray}
with $\mu_\Phi=51.2$ and $\sigma_\Phi =0.77$.
Certainly, in the low flux regime, a log-normal is not the best fit of the observed flux as there are
many BMRs with fluxes smaller than $5\times10^{21}$~Mx. However, smaller BMRs may not
contribute 
much net poloidal flux because of their smaller flux and large scattered tilts \citep{SK12}. 
The above distribution with $\Phi_0 = 1$, plotted in \Fig{fig:fluxdist}, is obtained from \cite{Mu15} 
based on observations of Solar and Heliospheric Observatory/Michelson Doppler Imager (SOHO/MDI) 
magnetograms during 1996--2010.
Different data sets in their publication produced slightly different 
values of $\mu_\Phi$ and $\sigma_\Phi$ \citep[see also][]{ZWL10,LCC15}.
Once the flux of the BMR is obtained from the above distribution, the radius is automatically set by specifying 
a fixed value for the surface field strength of 3~kG. As discussed in MT16,
if this radius turns out to be comparable or smaller than the grid size of the domain, 
then we set the radius 
at five times the grid size and the field strength is reduced accordingly.

The half distance between centers of two spots of a BMR is chosen to be $1.5$ times the radius of the spot.
As in our earlier model, we have assumed spots to be disconnected from
their parent spot-producing fields.  
The surface fields are extrapolated downward using a potential field approximation 
as described in MT16, which yields the full 3D structure of a BMR.
In our model, BMRs are assumed to be rather shallow by choosing the radial
field of the spots to be zero at $r_s = 0.9R$.

\begin{figure*}
\centering
\includegraphics[width=2.20\columnwidth]{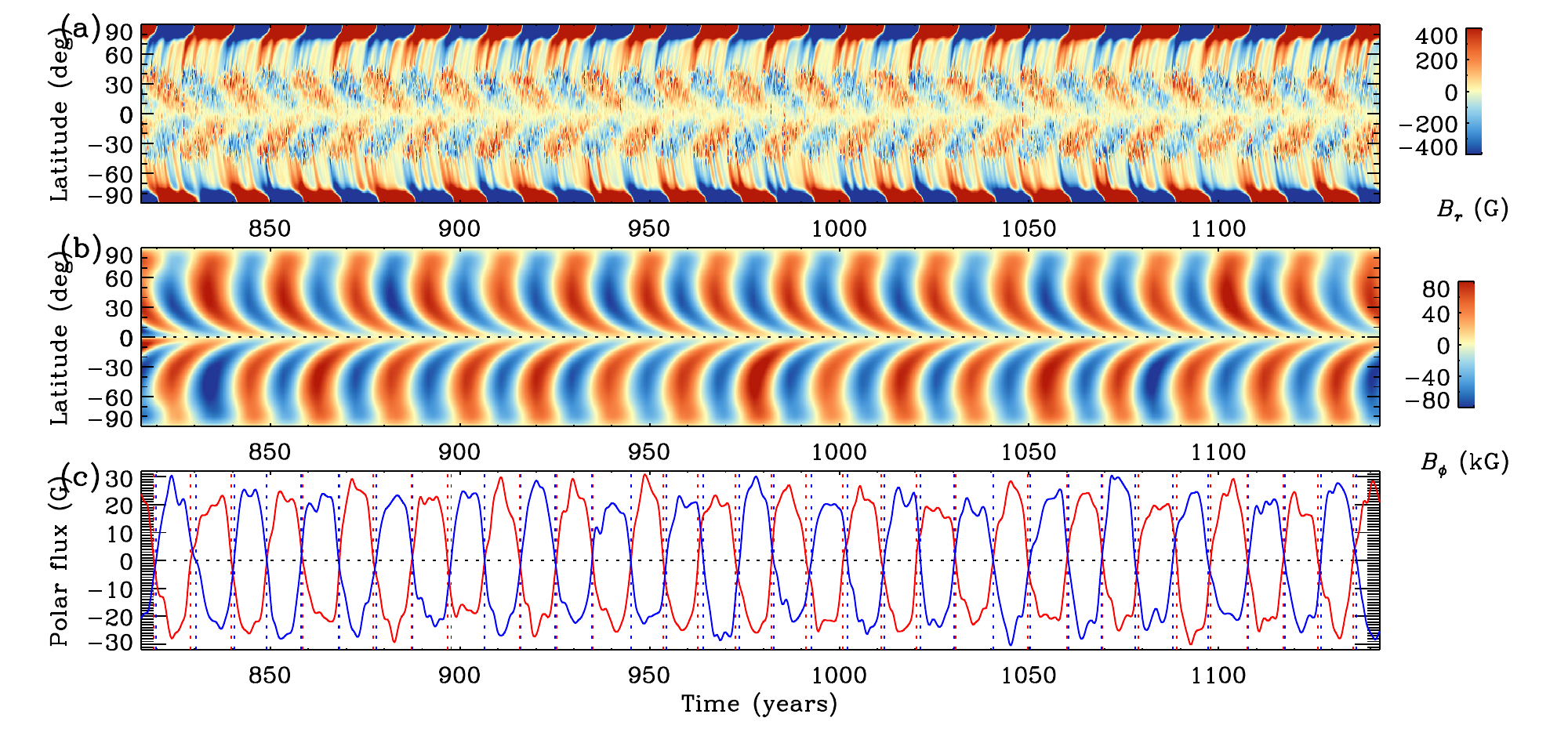}
\caption{
Results from Run~A6: temporal evolutions of (a) the radial field $\brac{B_r(R,\theta,\phi)}_\phi$, 
(b) the toroidal field $\brac{B_{\phi}(0.72R,\theta,\phi)}_\phi$, and
(c) the polar flux density, computed by averaging surface $\brac{B_r}_\phi$ from $75^\circ$ latitude
to the pole (red: north, blue: south). 
The vertical dashed lines
show the times of zeros of the mean polar flux.
Note that the color scales in (a) and (b) are saturated at $\pm400$~G and $\pm80$~kG, 
while the extrema are $[-5.8,5.9]$~kG, and $[-90,106]$~kG, respectively.
}
\label{fig:lowdifhr}
\end{figure*}

In our previous publications (MD14, MT16), we have used the standard Joy's law:
$\delta  = \delta_0  \cos\theta$ \citep{Hale19,Das10,SK12}
for tilt angles of BMRs. Here we make two modifications
in it. One is made by adding a random component $\delta_{\rm f}$
around Joy's law. In observations,
we notice that Joy's law is a statistical law and there
is a considerable scatter around it \citep{How91,SK12,MNL14,pavai15}.
Particularly, from the analysis of BMRs measured during 1976--2008, \citet{Wang15} reported
that the fluctuations of the tilts roughly follow a Gaussian distribution:
\begin{equation}
 f(\delta_{\rm f}) = \frac{1}{\sigma_\delta \sqrt{2\pi} } \exp[ -\delta_{\rm f}^2 / (2\sigma_\delta^2) ],
\end{equation}
with $\sigma_\delta \approx 15^\circ$.  We understand that a
Gaussian is not the best fit to the observed fluctuations of the
data because of its asymmetric shape and considerable outliers near
two ends of the distribution. However, to capture the broad picture
of the tilt fluctuations in our model, the above Gaussian
distribution is sufficient \citep[see also Figure~3 of][for the
distribution of BMR tilts within $15$--$20^\circ$ latitudes]{SK12}.
Another modification to Joy's law that we implement here is the tilt-angle saturation; 
the tilt is suppressed for strong progenitor toroidal fields. 
Thus the tilt used in our model is given by 
\begin{equation} 
\delta  = \frac{\delta_0 \cos\theta + \delta_{\rm f}} {1 + ( \hat{B}(\theta,\phi,t) /\Bsat)^2}, 
\label{eq:joys}
\end{equation} 
where $\delta_0 = 35^\circ$ and $\Bsat$ (the saturation field strength) is chosen to $1\times10^5$~G.  
In thin flux tube simulations, tilts of the BMRs are produced due
to the Coriolis force acting on the toroidal flux tubes during their rise in the CZ \citep[e.g.,][]{DC93,FFM94}. 
When the spot-producing toroidal field is strong, the field rises fast and the Coriolis force
does not get much time to tilt it. Thus, from this theoretical argument, we expect some quenching in the tilt.
In observations, we find some evidence of tilt quenching with the BMR flux \citep{Das10,SK12},
although the picture is less transparent due the lack of detailed analysis. In any case,
we shall explore whether the above magnetic field dependent
nonlinearity is sufficient to stabilize the growth of the magnetic
field in \Eq{eqind} and in the future work, we shall consider other possible saturation mechanisms.  
We note that in our previous model (MD14, MT16), 
dynamo saturation was implemented by saturating the flux content of BMRs rather than their tilt.  
The tilt angle saturation we use here has more physical justification.

For boundary conditions, we use radial field on the surface and perfect conductor at
the lower boundary. For the initial seed field, we use a weak dipolar magnetic field.

\begin{figure}
\centering
\includegraphics[width=1.0\columnwidth]{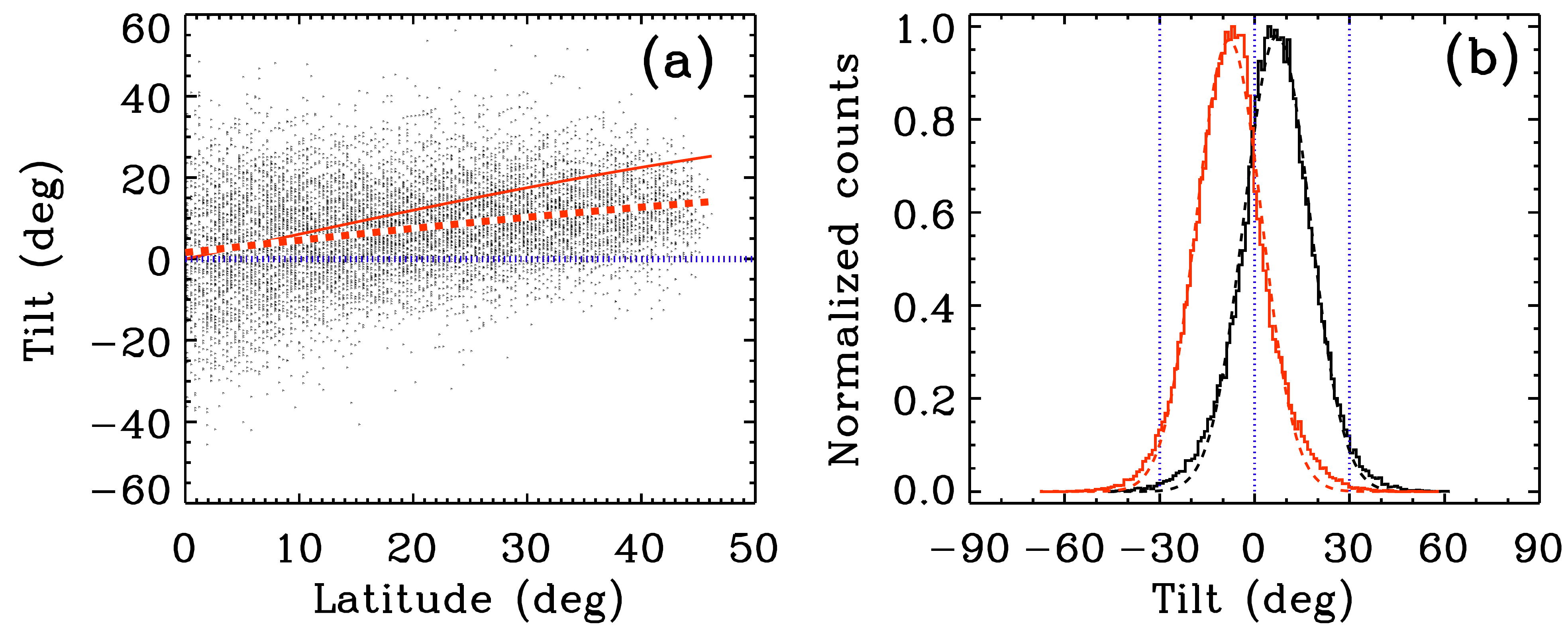}
\caption{
(a) BMR tilts versus latitudes for the northern hemisphere.
The solid, dashed, and dotted lines respectively show the actual Joy's law: $\delta  = \delta_0 \cos\theta$, 
the linear fit of tilts, and the zero line.
(b) Black and red are normalized histograms of tilts obtained from northern and southern hemispheres, respectively. 
Dashed lines are the Gaussian fits
with $\mu_\delta = 7^\circ$ and $8^\circ$,
and $\sigma_\delta = 10.4^\circ$ and $10.3^\circ$ (half width at half maximum HWHM $=12.3^\circ$ and $12.2^\circ$) for north and south, respectively. Vertical lines mark $\pm30^\circ$ and $0^\circ$ tilts.
These plots are obtained from
Run~A6 presented in \Fig{fig:lowdifhr}, except in panel (a) where data only from $850$ to $875$ years are used.
}
\label{fig:mod:tilt}
\end{figure}

\section{Results for fixed BMR delay distribution}
\label{sec:fixedTD}

As discussed in \Secs{sec:int}{sec:mod}, we consider two ways in which the photospheric flux is linked to the deep toroidal flux.  The first is by making the BMR flux proportional to the deep toroidal flux.  These runs are labeled with ``A" and discussed in this section.  The second is to fix the flux distribution and instead link the BMR emergence rate to the deep toroidal flux.  These runs are described in \Sec{sec:BdepTD}.

For the model with fixed delay distribution, we scale the observed BMR flux 
with the toroidal field at the base of the CZ, such that the 
BMR flux in the model, 
$\Phi_s = (\hat{B}(\theta_s,\phi_s,t) / \Bsat) \Phi$.
Here 
($\theta_s,\phi_s$) is the location of the BMR
and 
$\Phi$ is the BMR flux obtained from the observed distribution given in \Eq{eqflux}.
Using this BMR flux and other ingredients as specified in \Sec{sec:mod}, we run the dynamo model to
simulate the solar cycle.  However, when we use the observed BMR flux
distribution with $\Phi_0=1$, we get decaying solutions
for different parameters of the model.  Runs~A1--A2 in Table~1
represent these decaying solutions.  Boosting up the observed
flux distribution even by a small factor does not help. We realized
that when the flux distribution is increased at least by a factor of 28 (i.e., $\Phi_0=28$), 
we get a growing solution; see Runs~A3--A4.  
The sustained dynamo action is easier if we add a downward magnetic pumping ($\gamma_r$); compare Run~A3 with A5 and Run~A4 with A6. 
\Fig{fig:lowdifhr}
displays time evolutions of magnetic fields for about $300$~years from
 Run~A6, in which a surface magnetic pumping (\vpS) of $2$~\mps\ is used. 
It is apparent that the magnetic
field is stable and the overall cycle amplitude is limited in
time. The dynamo saturation mechanism is the
quenching of the tilt angle introduced through \Eq{eq:joys}.
Because of this quenching, the mean tilt, shown by the dashed
line in \Fig{fig:mod:tilt}(a), deviates from the actual Joy's
law:  $\delta  = \delta_0  \cos\theta$ (solid red line). 
We note that a recent coupled $2$D$\times2$D
\bl\ model of \citet{LC16} also produces a stable solution with the
tilt quenching.

\begin{table*}
\caption{Summary of the simulation runs.
}
\begin{center}
\begin{tabular}{lccccccccc}
\hline
Run &$\Phi_0$& $\sigma_\delta$ &$\eta_{\rm CZ}$ & \vpCZ,~\vpS&$\tilde{B}_{tor}$~~~~$\tilde{B}_{r}$& Parity of $B_\phi$ & Period & \# of BMRs & Variability of\\
     &    &       &(cm$^2$ s$^{-1}$)&(\mps)& (kG)~~~~(G)     &[$\overline{\rm SP}_\phi$ (${\sigma_{\rm SP}}_\phi$, T$_{\rm SP}$)]& (years) & (per cycle) & $\overline{B_r}$~~~~~SSN\\
\hline
A1   &  1 & $0^\circ$  &8.0$\times10^{10}$ & 0,~0&subcritical& -- & -- & -- & -- \\
A2   &  1 & $0^\circ$  &5.0$\times10^{10}$ & 0,~0&subcritical& -- & --  & -- & -- \\
A3   &  28& $0^\circ$  &8.0$\times10^{10}$ & 0,~0&~58~~~~~310&$-0.70$ (0.30, 123)&~9.0 &~3132&~~9\%~~~~~~--~~~~\\
A4   &  28& $15^\circ$ &8.0$\times10^{10}$ & 0,~0&~56~~~~~300&$-0.77$ (0.24, 125)&~8.9 &~3157& 11\%~~~~~--~~~~\\
%
A5   &  16& $0^\circ$ &8.0$\times10^{10}$ & 0,~2&~67~~~~~300&$-0.89$ (0.18, 141)&~9.4 &~3336& ~~9\%~~~~~~--~~~~\\
A6   &  16& $15^\circ$&8.0$\times10^{10}$ & 0,~2&~65~~~~~290&$-0.85$ (0.23, 143)&~9.6 &~3374& 11\%~~~~~--~~~~\\
A6$^\prime$&16& $15^\circ$&8.0$\times10^{10}$ & 0,~2&~56~~~~~270&$-0.70$ (0.29, 142)&~9.5 &~3188&10\%~~~~~--~~~~\\
\hline
AB1  & 16 & $15^\circ$&8.0$\times10^{10}$ & 0,~2&150~~~~1000&$+0.82$ (0.20, ~15)&~7.6 &33371&24\%~~~~~4\%\\
\cdashline{1-10}
B1   &  1 & $15^\circ$&$1.5\times10^{12}$&0,~0&subcritical& -- &-- & -- & --\\
B2   &170 & $15^\circ$&$1.5\times10^{12}$&0,~0&390~~~~8300&$-0.92$ (0.20, 165)& ~5.2 & 11981&14\%~~~~~2\%\\
B3   &  1 & $15^\circ$&$1.1\times10^{12}$&0,~15&subcritical&-- &-- & -- & --\\
%
B4   &  1 & $15^\circ$&$1.5\times10^{12}$&4,~20&subcritical&-- &-- & -- & --\\
B5   &  1 & $15^\circ$&$1.0\times10^{11}$&2,~20&~11~~~~~~12&$-0.67$ (0.32, 296) &13.2 &~~568&52\%~~~~59\%\\
%
B6   &1.5 & $15^\circ$&$1.5\times10^{12}$& 2,~35&8.5~~~~~~37&$-0.96$ (0.13, 243)&14.9 & ~1343&12\%~~~~19\%\\
B7   &1.3 & $15^\circ$&$1.5\times10^{12}$& 5,~25&6.8~~~~~~21&$-0.96$ (0.10, 427)&11.2&~~994&18\%~~~~28\%\\
B8   &1.5 & $15^\circ$&$1.5\times10^{12}$&10,~20&~12~~~~~~46&$-0.89$ (0.21, 317)&~7.3 & ~2420 &40\%~~~~42\%\\
%
B9  &2.4 & $0^\circ$  &$1.5\times10^{12}$&2,~20&~22~~~~~140&$-0.98$ (0.07, 1092)&10.5 & ~3947 & 11\%~~~~14\%\\
B10  &2.4 & $15^\circ$&$1.5\times10^{12}$&2,~20&~30~~~~~190&$-0.96$ (0.10, 1465)&10.5 & ~6052 & 35\%~~~~41\%\\
B11 &2.4 & $30^\circ$ &$1.5\times10^{12}$&2,~20&~31~~~~~185&$-0.90$ (0.16, 4262)&10.8 & ~6119 & 46\%~~~~54\%\\
B12 &3.4 & $15^\circ$ &$1.5\times10^{12}$&2,~20&~38~~~~~190&$-0.97$ (0.08, 1162)&12.8 & ~10817& 17\%~~~~26\%\\
B13 &2.4 & $15^\circ$ &$1.5\times10^{12}$&2,~20&~21~~~~~160&$-0.94$ (0.11, 734)& ~9.3 & ~4366 & 23\%~~~~30\%\\
\cdashline{1-10}
C1  &2.4 & $15^\circ$ &$1.5\times10^{12}$&2,~20&~26~~~~~170&$-0.94$ (0.12, 493)&~9.9 & ~4641 & 26\%~~~~29\%\\
\cdashline{1-10}
D1  &3.0 & $15^\circ$ &$1.5\times10^{12}$&2,~20&~14~~~~~~64&$-0.90$ (0.16, 1339)&13.9 & ~1190 & 22\%~~~~31\%\\
\hline
\end{tabular}
\end{center}
\tablecomments{
In the A series of simulations, the delay distribution of BMR eruptions is fixed (i.e., fixed $\taus$ and $\taup$)
but the observed flux distribution is scaled by the toroidal field at the base of the CZ,
while in all other simulations the delay distribution is dependent on the magnetic field through \Eq{eqtau}
but the flux distribution is fixed. 
Runs B12, B13, C1, and D1 are the same as Run B10, except in Run~B12 $\Bsat$ is four times smaller and $\Phi_0=3.4$,
in Run~B13 $\etat$ in the tachocline is same as that in the CZ, in Run~C1 the quenching is in the BMR flux and not in the tilt, 
and in Run~D1 different forms of magnetic field dependent $\taus$ and $\taup$ (see text).
The root-mean-square (rms) values of the mean toroidal and poloidal fields over 
the entire computational domain are denoted by $\tilde{B}_{tor}$ and $\tilde{B_{r}}$, respectively.
Symbols $\overline{\rm SP}_\phi$ and ${\sigma_{\rm SP}}_\phi$ respectively denote
the mean and the standard deviation of the parity of 
$\brac{B_\phi(0.72R,\theta,\phi)}_\phi$ 
computed over T$_{\rm SP}$~years of data.
Periods are computed from the power spectrums of the azimuthal averaged toroidal field at $r=0.72R$, 
integrated over $0$--$30^\circ$ latitudes for the A series of runs
or from the yearly averaged sunspot number (SSN) for all other runs. 
In last two columns, the variabilities of the peak polar field (${\overline{B}_r}$) and the peak SSN are measured as 
$\sqrt{ \frac{1}{N}\sum_{i=1}^N (Q_i - \overline{Q})^2 }/\overline{Q} \times 100 \%$, 
where $Q =$ peak ${\overline{B}_r}$ or peak SSN, $N=$ total number of cycles, and $\overline{Q} = \frac{1}{N}\sum_{i=1}^N Q_i$.
Runs~A3--A6 and AB1 have spatial resolutions of $340\times512\times1024$ in 
$r$, $\theta$, and $\phi$, respectively, 
while all other runs, 
including A6$^\prime$,
have resolutions of $200\times256\times512$.
}
\label{table1}
\end{table*}

\begin{figure}
\centering
\includegraphics[width=1.0\columnwidth]{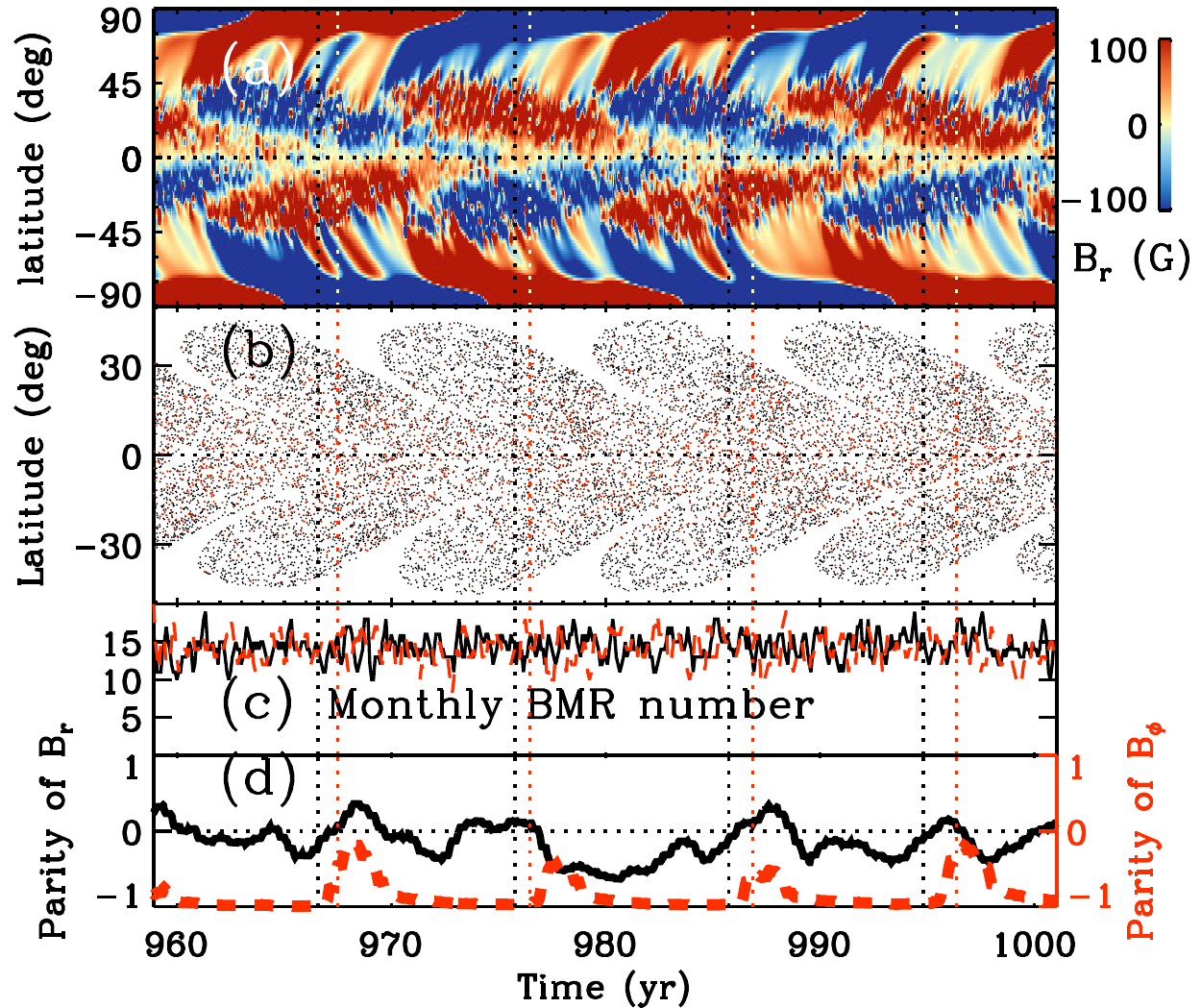}
\caption{
Temporal variations of (a) $\brac{B_r(R,\theta,\phi)}_\phi$, 
(b) BMR latitudes, (c) monthly number of BMRs (solid/dashed: north/south), 
(d) ${\rm SP}_r$ and ${\rm SP}_\phi$ (dashed)
from Run~A6 presented in \Fig{fig:lowdifhr} but shown only for a few cycles.
In panel (b), the red color shows BMRs with wrong tilts violating the sense of Joy's law, i.e., negative in the northern hemisphere and positive in the southern hemisphere.
Vertical dotted lines show times of reversals of the low-latitude bottom toroidal flux; black/red: north/south.
Note that the color scale in (a) is saturated at $\pm100$~G, while the maximum field strength is $\pm5.5$~kG.
}
\label{fig:LDzoomed}
\end{figure}

From the butterfly diagrams in \Fig{fig:lowdifhr}, we 
recognize that the magnetic field is largely dipolar as both the toroidal and radial fields are asymmetric across the equator.
However, to make a quantitative measure of the equatorial symmetry of different components of the magnetic field,
we compute the symmetric parity (SP) by cross correlating the fields between two hemispheres in the same way as done in \citet{CNC04}, i.e.,
\begin{equation}
{\rm SP}_j (r,\theta,t)  = \frac{ \int_{t-\frac{T}{2}}^{t+\frac{T}{2}} \left (B^N_j  - \overline{B^N_j} \right) \left(B^S_j  - \overline{B^S_j} \right) {\rm d}t^\prime} {\sqrt{ \int_{t-\frac{T}{2}}^{t+\frac{T}{2}}\left(B^N_j  - \overline{B^N_j} \right)^2 {\rm d}t^\prime \int_{t-\frac{T}{2}}^{t+\frac{T}{2}} \left(B^S_j  - \overline{B^S_j} \right)^2 } {\rm d}t^\prime},
\label{eq:parity}
\end{equation}
where $j$ stands for $r$, $\theta$ or $\phi$ component, 
$B^N_j = \brac{B_{j}(r,\theta,\phi,t^\prime)}_\phi$, $B^S_j = \brac{B_j (r,\pi-\theta,\phi,t^\prime)}_\phi$,
and overlines denote the average over period $T$.
To identify the short-term temporal variation of the parity, we take $T=3.73$~years.
In all the cases, we compute the parity at a fixed radius (at $0.72R$ for $B_\phi$ and $R$ for $B_r$) 
and average over latitudes ($\pi/2 < \theta \le \pi$).
From the above definition of parity, we expect, ${\rm SP}_j=1$ for a perfect symmetric field
and $-1$
for an antisymmetric field.
We note that for a dipolar field, ${\rm SP}_r = -1$, 
${\rm SP}_\theta = 1$ and ${\rm SP}_\phi = -1$ and the reverse is true for the quadrupolar field.

On taking the toroidal field at $r=0.72R$ and the radial field at $r=R$ from Run~A6, we compute the mean parity
of toroidal field $\rm SP_\phi (t)$ and the mean parity of radial field ${\rm SP}_r (t)$.
These quantities are displayed in \Fig{fig:LDzoomed}(d) for a few cycles.
We observe that the parity of the bottom toroidal field is more antisymmetric
than that of the surface radial field. The latter is largely deviated from $-1$ mode due to
continuous BMR eruptions at low latitudes. Thus if we had computed the parity of high latitudes $B_r$,
then we would have obtained the value close to $-1$ (dipolar).
When we compute the average parity over the whole
simulation run, we obtain $\overline{\rm SP}_r = -0.11$ and $\overline{\rm SP}_\phi = -0.85$. 
The respective standard deviations of these parities are $0.23$ and $0.22$, suggesting that they
have considerable deviations from their antisymmetric modes. These are seen in \Fig{fig:LDzoomed}(d)
that parities tend toward the
symmetric (quadrupolar) mode during solar maxima when new BMRs emerge on the surface.
Then the decay of these BMRs produces largely antisymmetric (dipolar) field at the solar minima.
This is broadly consistent with observations \citep{DBH12}.

Returning to \Fig{fig:lowdifhr}, we notice that this simulation 
also produces polarity reversals with an average period of $9.6$~years, 
equatorward migration
of toroidal field at low latitudes, and poleward migration of radial
field, all broadly consistent with observations. 

As given by \Eq{eq:joys}, the tilt angle in this model
has a random component following a Gaussian distribution with
$\sigma_\delta=15^{\circ}$ around Joy's law.
Because of this random component, the actual tilt angle in our model
has a considerable variation.
Since the poloidal field generated by the BL mechanism depends sensitively on the BMR tilt \citep{Das10,JCS14,HCM17}, 
its random scatter gives rise to cycle variability. 
This greatly enhances the relatively modest cycle variability arising just from the random time delay (MT16).

The variation of tilt angle has an even larger effect when the tilt acquires a ``wrong" sign,
i.e., negative in the northern hemisphere and positive in the southern hemisphere.
The word ``wrong'' here is not intended as a value judgment.  Rather, the ``right" sign of a tilt is defined by Joy's law.  The random fluctuations can lead to a tilt that violates Joy's law.  This is the sense with which it is ``wrong".   We note that having a wrong tilt does not necessary imply that the BMR violates Hale's polarity rule.
Wrong tilts happen frequently in our model as seen in \Fig{fig:mod:tilt} or \Fig{fig:LDzoomed}(b) and
produce a poloidal field of 
the opposite polarity.
This is reflected in \Fig{fig:lowdifhr}(a) 
and more clearly in \Fig{fig:LDzoomed}(a), 
where we notice mixed polarity field and frequent
polar surges of opposite polarity. This type of mixed polar
field is frequently found in observations; see e.g., Figure 8a of \citet{Scott15}.
The radial polar flux density (flux per unit area) 
as shown in \Fig{fig:lowdifhr}(c) has a considerable cycle to cycle variation.
The amount of variation 
$\sqrt{ \frac{1}{N}\sum_{i=1}^N ({\overline{B}_r}_i - \overline{B}_r^{\rm avg}})^2 /\overline{B}_r^{\rm avg} \times 100 \% = 12.2\%$ 
(where $\overline{B}_r$ is the peak value of the radial flux density computed by averaging over $15^\circ$ around the pole, 
$\overline{B}_r^{\rm avg} = \frac{1}{N}\sum_{i=1}^N {\overline{B}_r}_i$, 
and $N = 34$, the total number of cycles).

The mean polar field in \Fig{fig:lowdifhr}(a) is larger than 
the observed value\footnote{see http://solarscience.msfc.nasa.gov/images/magbfly.jpg}, 
although the observed polar field
is not reliably measured (because of the resolution limit and the
projection effect). However, when we measure the
mean polar flux density in high latitudes, say from $75^\circ$ latitude
to the pole as shown in \Fig{fig:lowdifhr}(c), we obtain a strength of 
the mean polar field around $20$~G which is close to the observed range.
Another discrepancy between the present model and the observation is
that a significant overlap between two cycles at each minimum; the cycle
starts much before the end of its previous cycle \citep[compare our
\Fig{fig:lowdifhr}{a} with Figure 8a of][]{Scott15}.  
The discrepancy can be attributed to our incomplete understanding of flux emergence and how to parameterize it with SpotMaker.
Nevertheless,
the overall morphology of our radial field resembles observations more closely
than our previous model without tilt
fluctuations (see MT16, for example) 
and also previous 2D dynamo models \citep{Cha10,Kar14a}.

We have demonstrated that in the present model, the cumulative effect of the short term 
variations of the tilt angle is capable of producing
a variation in the magnetic cycle as seen in \Fig{fig:lowdifhr}(c). 
Thus we can conclude that a potential cause of solar cycle variability is the observed scatter of the tilt angle \citep{SK12,Wang15,pavai15}.
While we in our 3D dynamo model and \citet{LC16} in their coupled $2$D$\times2$D model explicitly demonstrate this,
the original idea was known since the work of \citet{CD00}. 
Recently, \citet{Ca13} demonstrated this idea using observations, 
while \citet{JCS14} for the first time quantified the effect of the tilt scatter on the polar field using a surface flux transport model.
Based on this idea 
many authors \citep[e.g.,][]{CK09,YNM08,OK13} modeled irregular features of the solar cycle by including fluctuations in the \bl\ $\alpha$ term
of their 2D flux transport dynamo models.  

We mention that fluctuations of BMR tilts in
our model were approximated by a Gaussian distribution with
$\sigma_\delta=15^{\circ}$. In observations, however, there is large
scatter near the two tails of distribution which is not captured in 
our Gaussian model; compare our \Fig{fig:mod:tilt}(b) with Figure~2 of \citet{Wang15} or Figure~12 of \citet{pavai15}.
Thus in our model, if we had considered the tilt angles from the actual 
observations, then we would have achieved
even more variation in the magnetic field
than we have obtained here.

\begin{figure}
\centering
\includegraphics[width=1.0\columnwidth]{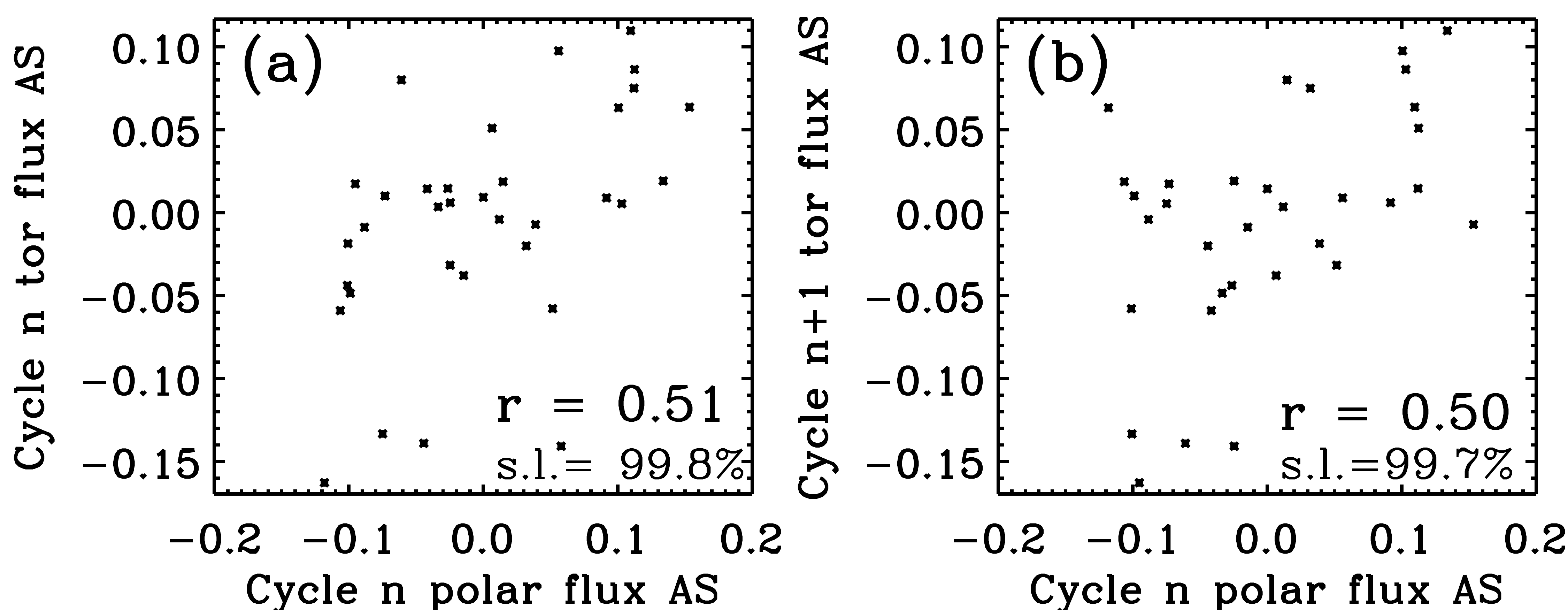}
\caption{Results from Run~A6: scatter plots between the asymmetry of the peak surface polar flux 
and the asymmetry of the peak toroidal flux at the base of the CZ
for (a) the same cycle, and (b) for the next cycle.
}
\label{fig:lowdifasy}
\end{figure}

Though the dynamo maintains a strong hemispheric coupling,
it also exhibits a noticeable hemispheric asymmetry (\Fig{fig:lowdifhr}(c)). 
Thus we find nonzero values for the asymmetry in peak surface polar fluxes, 
as measured by
AS$_{\rm pol} = (|\overline{B}_r^{\rm N}| - |\overline{B}_r^{\rm S}|)
/ (|\overline{B}_r^{\rm N}| + |\overline{B}_r^{\rm S}|$). 
We note that this is not the parity of the polar field computed in \Eq{eq:parity}.
If there were no asymmetry introduced in the poloidal flux 
generation, then the asymmetry in the toroidal flux would be reflected 
in the poloidal flux and we would have obtained a strong correlation 
between these two.
Nonetheless, we find only a moderate
correlation (with 
linear Pearson correlation 
coefficient r $=0.51$ with a significance level ($(1-p)\times100\%$) of $99.8$\%)
between the polar flux asymmetry and the low-latitudes toroidal flux asymmetry (AS$_{\rm tor}$); see \Fig{fig:lowdifasy}(a).
This suggests that the asymmetry in the toroidal flux is not the only
cause of the asymmetry in the poloidal flux, rather it can be produced from the 
asynchronous BMR emergence rate and the tilt angle. The
asymmetric polar flux should eventually cause an asymmetry in
the toroidal flux.  However due to hemispheric coupling at the
equator, the asymmetry gets reduced over the time and we find a
moderate correlation between the asymmetry in polar flux and the
asymmetry in the next cycle toroidal flux, as shown in
\Fig{fig:lowdifasy}(b).  

\begin{figure}
\centering
\includegraphics[width=1.0\columnwidth]{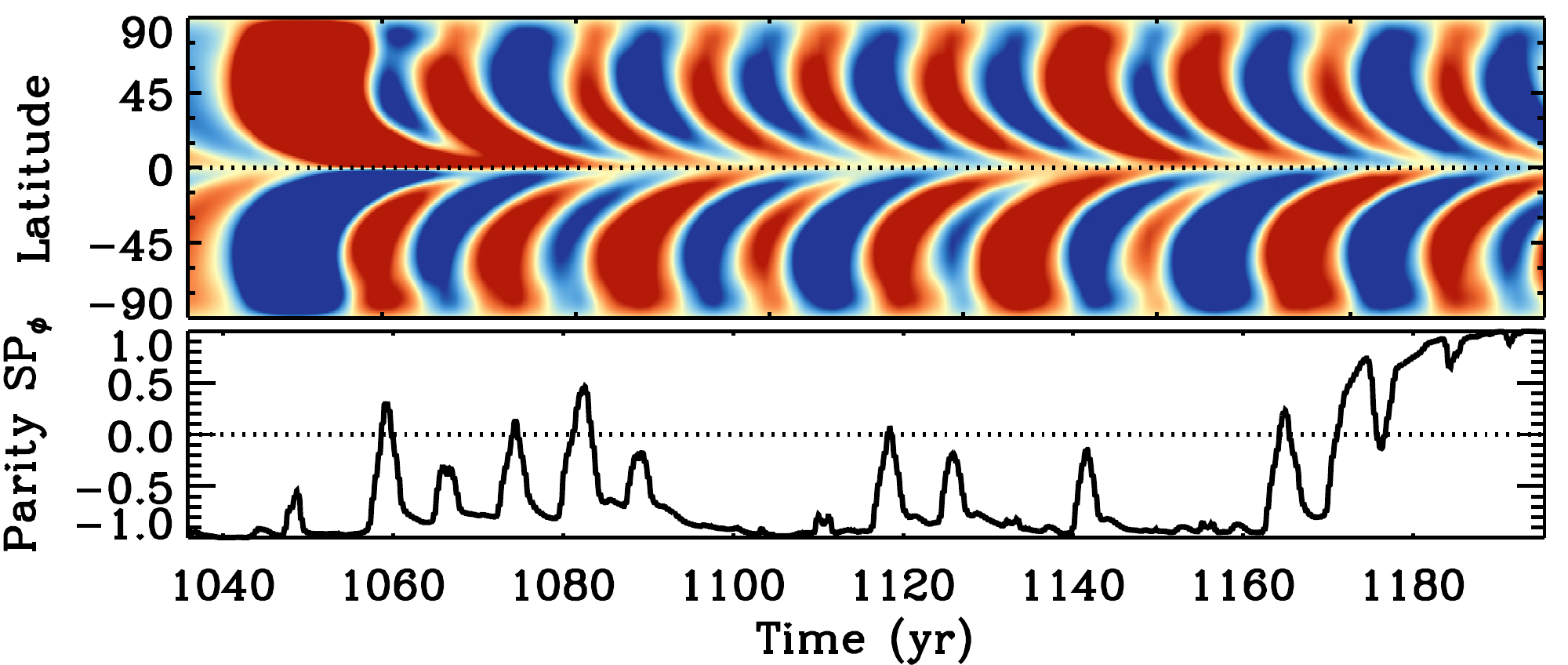}
\caption{
The temporal evolutions of $\brac{B_\phi(0.72R,\theta,\phi)}_\phi$ (top) 
and the parity of this field (bottom), obtained from Run~AB1 which was
started from the 
dipolar field of Run~A6 and eventually settled to a quadrupolar solution.}
\label{fig:AB1}
\end{figure}

As discussed in \Sec{sec:mod}, when we do not have sufficient spatial resolution,
the sizes of the smallest BMRs are limited by the spatial grid size.
In Run~A6, the minimum size of BMRs is about 6.8~Mm. The number of BMRs below this size is very small and thus the net flux from these small BMRs
are negligible in the poloidal field generation.
Hence, the spatial resolution of this simulation ($340\times512\times1024$) is sufficient to capture the observed BMR spectrum.
However, when we reduce the resolution to $200\times256\times512$, then the sizes of the smallest BMRs are about 13.6~Mm.
Thus this resolution is not adequate to resolve the full BMR spectrum and therefore we find a noticeable difference in the dynamo solution; see Run~A6$^\prime$ in \Tab{table1} for this simulation. Although the morphology of the magnetic fields (not shown) are not too different
in comparison to Run~A6, we find considerably smaller values of the magnetic fields.  
The reason for the weaker field could be the following. In comparison to Run~A6, in Run~A6$^\prime$ the sizes of the smallest BMRs are larger but the BMR field strengths are smaller. Thus in Run~A6$^\prime$, most of the flux from these smallest BMRs gets easily canceled out and less flux is able to reach to higher latitudes. 
This causes weaker magnetic field in Run~A6$^\prime$.
Furthermore, values of $\overline{\rm SP}_r$ and $\overline{\rm SP}_\phi$ are different ($-0.14$ and $-0.70$ respectively are the values, in comparison to $-0.11$ and $-0.85$ for Run~A6).
Thus the parity of the dynamo solution is slightly sensitive to how we resolve the small BMRs.

\section{Cycle-Dependent BMR Emergence Rate}
\label{sec:BdepTD}

In calculations presented in \Sec{sec:fixedTD}, the time delay is computed from a log-normal distribution
given by \Eq{eq:delaypdf} with fixed $\taus$ and $\taup$. 
Hence as long as the spot-producing toroidal field exceeds the threshold field 
strength, the eruption can happen 
almost equally over the whole cycle. 
This contributes to the significant overlap between successive cycles as seen in \Fig{fig:LDzoomed}.
Well before the end of a cycle, emergences
from the next cycle start and we do not observe noticeable cyclic variation
in the BMR number; \Fig{fig:LDzoomed}(c). 
One potential cause of this problem is that we have chosen a fixed time delay distribution over the entire cycle
which is unlikely to be true. In observations, we find more BMRs during solar maxima than minima; 
see the thick and the thin solid lines in \Fig{fig:tdelay}.
From this data, we estimate that during solar minimum, the mean time delay $\taus$ 
(and mode $\taup$)
of BMR appearance is about 10~days (and 1~day). However as we go towards 
solar maximum, the emergence becomes more frequent and the mean time delay can be as short as a day.
Motivated by this observed feature, we make $\taup$ and $\taus$ as the toroidal magnetic energy
dependent such that in the northern hemisphere:
\begin{eqnarray}
\taup = \frac{ 2.2 ~ \mathrm{days}} { 1 + (B_b^N/ B_\tau)^2 }~ ,
\quad
\taus = \frac{ 20 ~ \mathrm{days}} { 1 +(B_b^N / B_\tau)^2 } ~,
\label{eqtau}
\end{eqnarray}
where $B_b^N$ is the azimuthal averaged toroidal magnetic field in a thin layer 
from $r=0.715R$ to $0.73R$ around $15^\circ$ latitudes and 
the value of $B_\tau$ is tuned to 400~G such that we get roughly same number of BMRs 
as in observations. 
For the southern hemisphere, we have the same expressions for $\taus$ and $\taup$, 
relating to the toroidal field in that hemisphere. In this way, no hemispheric
synchronization is made in the waiting time of the BMR appearance, which is physical.
We note that \citet{LC16} also used a magnetic field dependent delay in the BMR emergence 
through an emergence function, although their number of new BMRs at every numerical time step 
is extracted from a uniform distribution; see their Section 2.4.2.

We repeat the previous simulation, Run~A6, using the delay distribution 
with modified $\taus$ and $\taup$ as given in \Eq{eqtau} and no other changes.
This new simulation is labeled as Run~AB1 in \Tab{table1}
and the result is displayed in \Fig{fig:AB1}. 
The most distinct result we find from this simulation is that the initial dipolar
field is flipped to a quadrupolar field
in about $150$~years. The mean parity of bottom toroidal field over the whole simulation becomes $-0.46$, 
while for the last $15$~years it is $+0.82$. 
Thus when we make the BMR delay dependent on the magnetic field,
the quadrupolar mode is preferred over the dipolar mode.  
This suggests that both the dipolar and quadrupolar modes have comparable growth rates 
and both modes can readily be excited with relatively minor changes in the simulation parameters. 
We return to this issue in \Sec{sec:highdif} below.
Another point to note
in \Fig{fig:AB1} is that the overall dynamo efficiency is larger and the cycle period is shorter 
than the previous case of a fixed delay distribution. The reason is not difficult to understand.
Once the toroidal field at the base of CZ is stronger, it reduces $\taus$ and $\taup$
to make the BMR eruption more frequent. This frequent eruption
makes the poloidal field production faster, which ultimately causes
the stronger fields and faster polarity reversals.

\subsection{Diffusion-Dominated Regime}
\label{sec:highdif}
\subsubsection{Steady dynamo solution}
We recall that in the previous models, the bulk diffusivity $\eta_{\rm
CZ}$ was taken to be $8\times10^{10}$~\cmss, which is much smaller
than the surface diffusivity ($\eta_{\rm S} = 3\times10^{12}$~\cmss).
Previous studies from 2D \bl\ models have demonstrated that a weaker
diffusion promotes quadrupolar parity \citep{DG01,CNC04,HY10}.  Thus,
we increase $\eta_{\rm CZ}$ to a much larger value of $\sim10^{12}$~\cmss.
Unfortunately, at this higher value of $\eta_{\rm CZ}$, we do get
a decaying solution (Run~B1). One way to get a stable solution is to
shift the observed flux distribution toward larger values (i.e., $\Phi_0>1$).  
The cycle period then becomes unrealistically short; see Run~B2 in Table~1.

\citet{KC16} have shown that a downward pumping near the surface
reduces the diffusion of the flux across the surface and helps to
achieve a dynamo at a higher value of $\etat$ than hitherto.
However, even with a reasonable amount of surface pumping, we tend to get
decaying solutions unless we increase the observed flux distribution by a small value; 
see Runs~B3--B4. Obviously, the dynamo is efficient if we reduce $\eta_{\rm CZ}$ significantly; see Run~B5.
Thus by increasing the observed flux distribution by a small amount and 
using a surface pumping of about $20$~\mps, we get growing solutions 
for $\eta_{\rm CZ} > 1\times10^{12}$~\cmss; see Runs~B6--B11.
 
Comparing Runs~B2, B6, and B10, we notice that the cycle period increases 
with the increase of surface pumping \vpS. This is expected from the study of \citet{KC16}
that the pumping makes the dynamo efficient and thus allows us to use a smaller value of $\Phi_0$.
This makes the period longer by regulating the strength of the BL $\alpha$ effect.
However when the pumping in the whole CZ is increased, 
the downward transport of poloidal field becomes more efficient, 
reducing the time lag between poloidal and toroidal field conversion.
That is the reason for getting a shorter period at 
a stronger \vpCZ\ in Run~B8.

\begin{figure}
\centering
\includegraphics[width=1.00\columnwidth]{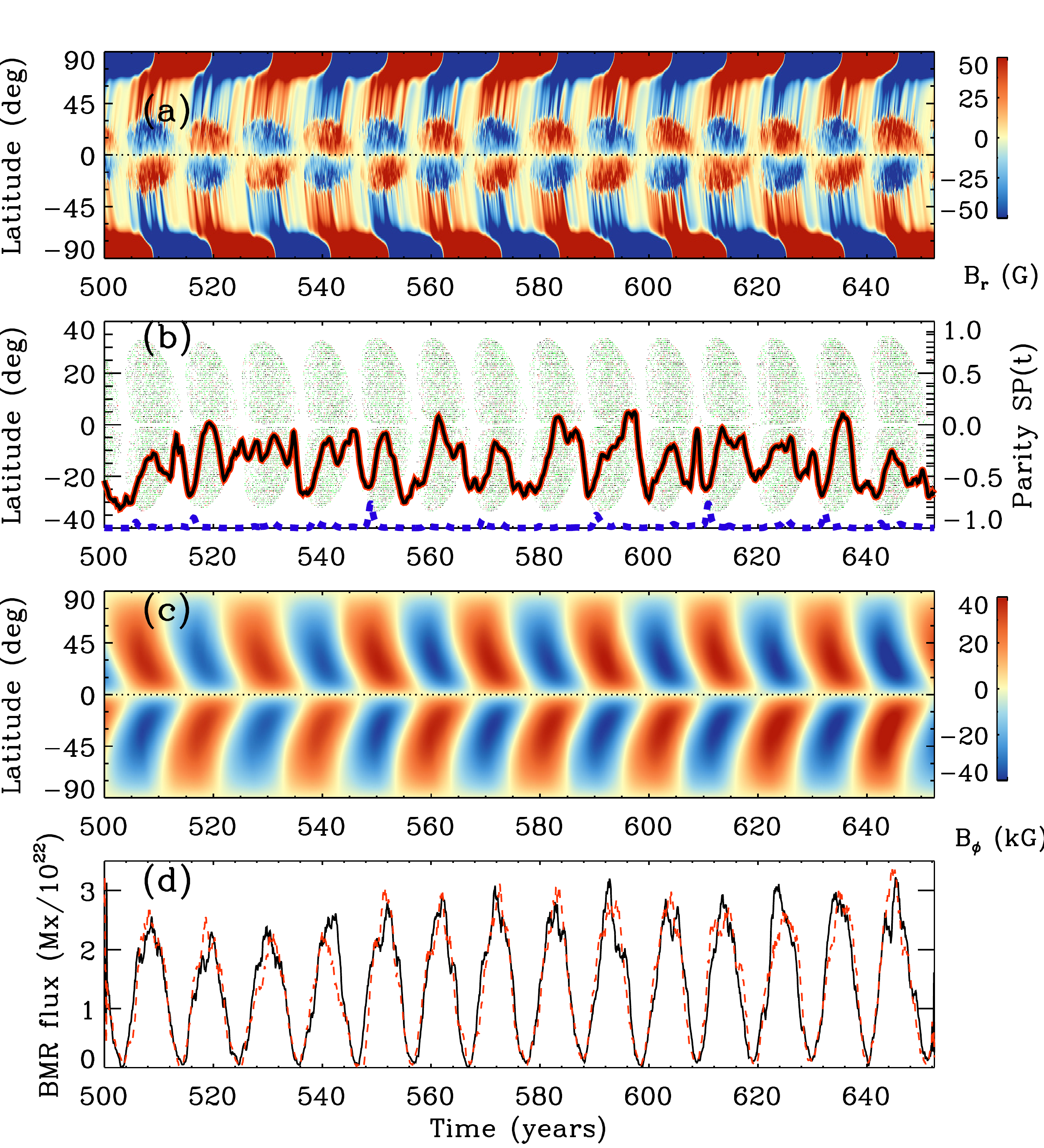}
\caption{Results from Run B9: temporal variations of (a) $\brac{B_r(R,\theta,\phi)}_\phi$, 
(b) latitudes of BMRs, (c) $\brac{B_\phi(0.72R,\theta,\phi)}_\phi$, and (d) daily BMR fluxes 
in Mx/$10^{22}$ (black/red: north/south) 
produced by this model.
In (b), points with different colors represent different sizes of BMRs. Green, black, and red
correspond BMRs of areas $< 500$ MHem (millionth of a solar hemisphere), $500$ MHem $\le$ areas $< 1000$ MHem, 
and  areas $\ge 1000$ MHem, respectively.
In the same panel, black/red and blue/dashed lines show parities, SP(t) computed over the four years of 
surface $B_r$ and the bottom $B_\phi$, respectively.
}
\label{fig:hdref}
\end{figure}

Results from Run B9
with $\eta_{\rm CZ} = 1.5\times10^{12}$~\cmss\ and with no fluctuations
around Joy's law are shown in \Fig{fig:hdref}. 
We note that in
addition to changes in $\eta_{\rm CZ}$, \vpCZ\ and \vpS, 
two more changes have been made in this B series of simulations and Runs~C1--D1.
First,
the meridional circulation profile has also been changed.
To enhance the efficiency of the toroidal flux advection in this diffusion-dominated model, 
we made the meridional flow speed near the base of the CZ faster than in the previous 
advection-dominated model (Runs~A1--A6 and AB1).  The latitudinal
component of this flow is shown by the solid line in \Fig{fig:mc}(a).
This new meridional flow is produced from the same analytical profile
as used in the previous advection-dominated model, which is the same as in
\citet{KC16}, except the prefactor, $(r - R_p)$ in their Equation 5 for the stream function
is removed and the surface flow speed is adjusted to $20$~\mps. 
Second, the spot-producing toroidal flux is computed in the tachocline, 
i.e., $r_a=0.7R$, and $r_b=0.715R$ are taken in \Eq{eqspotprodflc}.

Interestingly, in this diffusion dominated model, the mean parity of the bottom toroidal field, 
$\overline{\rm SP}_\phi$ is $-0.98$. 
Thus the toroidal field is largely 
antisymmetric across equator
(\Fig{fig:hdref})
with minimal variation in the parity (with standard deviation of 
$\rm SP_\phi = 0.07$).
However, the mean parity of $B_r(R,t)$, shown by the black/red line in \Fig{fig:hdref}(b), 
deviates most strongly from the 
antisymmetric (dipolar) mode
 during cycle maxima.
This is consistent with the analysis of solar data by \cite{DBH12},
namely that the parity of the observed radial magnetic field is dipolar during solar minimum, but 
becomes quadrupolar during solar maximum due to the emergence of many BMRs.

\begin{figure}
\centering
\includegraphics[width=0.55\columnwidth]{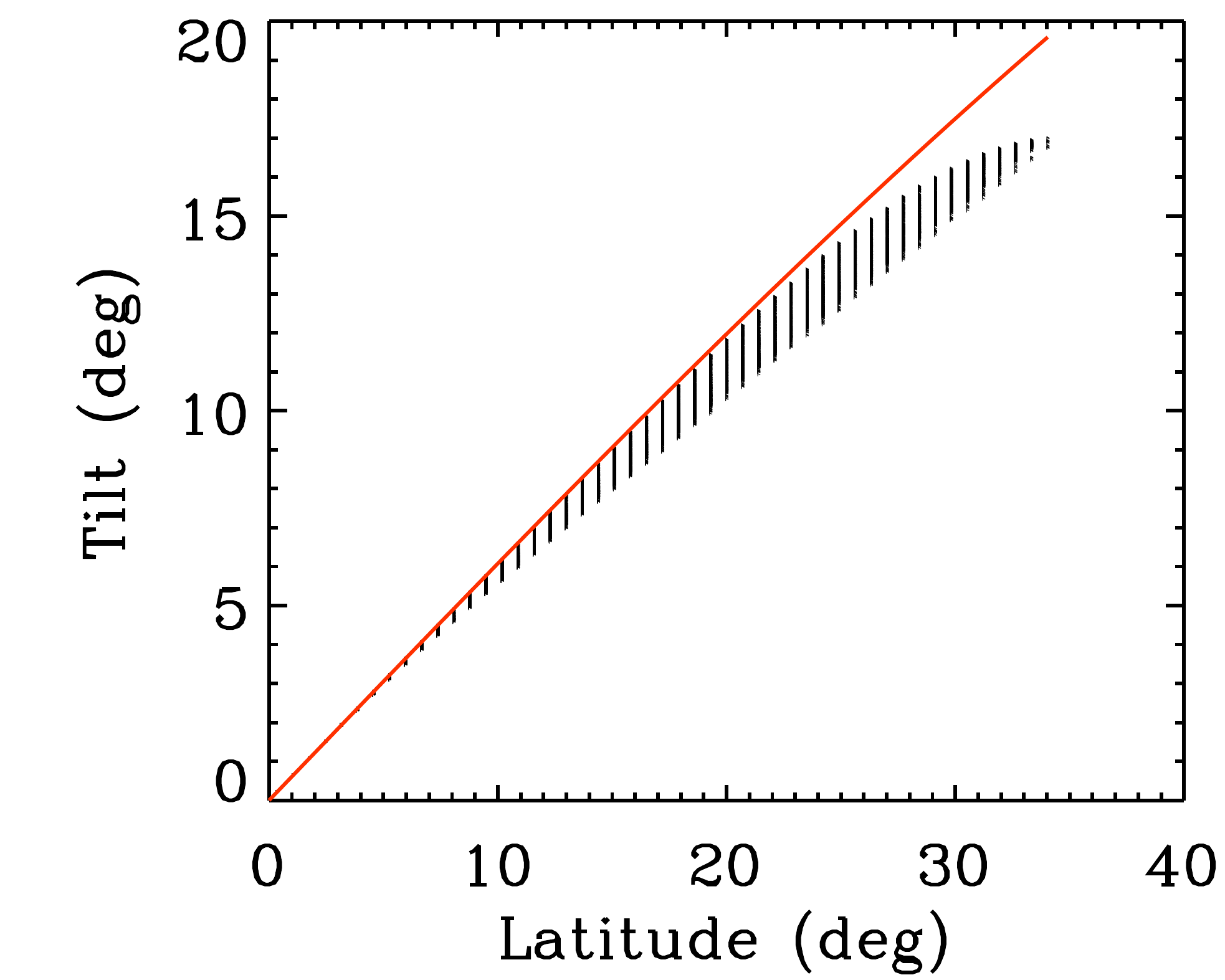}
\caption{Tilts of the northern hemisphere BMRs from Run~B9 as function of latitudes. The red line represents
the standard Joy's law: $\delta  = \delta_0  \cos\theta$.
}
\label{fig:hdtilt}
\end{figure}

Our simulation also produces most of the other features of the solar cycle.
However, there are some differences seen in this simulation compared to the previous advection-dominated model.
The tilt quenching, which produces the stable solution, is much weaker than
in the previous model; see \Fig{fig:hdtilt}.
This little quenching is sufficient to halt the dynamo growth.
Thus, the observational signature of tilt-angle quenching may be subtle and 
the weak evidence in favor of it \citep{Das10,SK12} 
may be sufficient to rank this as a viable candidate for dynamo saturation.

\begin{figure*}
\centering
\includegraphics[width=2.00\columnwidth]{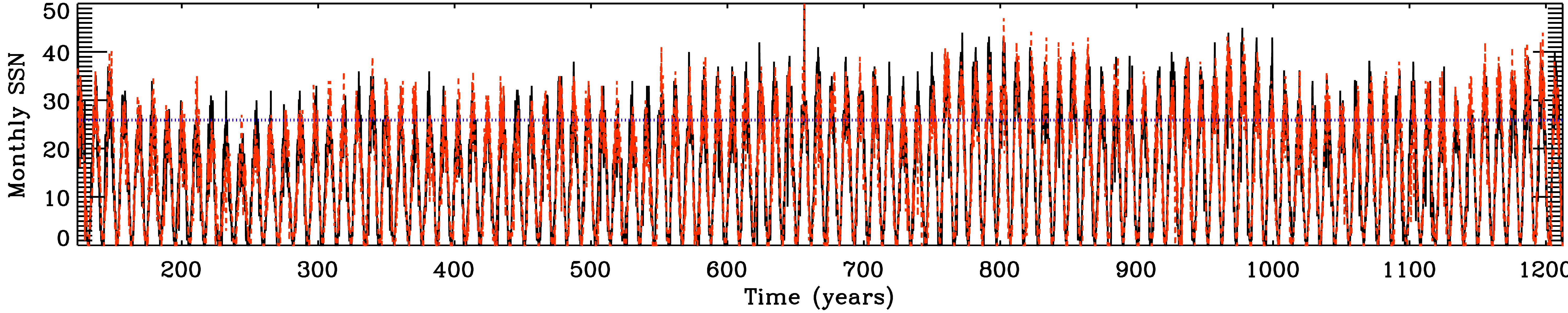}
\caption{Time series of the BMR number from the simulation presented in \Fig{fig:hdref} 
but highlighting a longer time interval.
The horizontal line shows the mean of peaks of the monthly BMRs obtained for last $13$ observed solar cycles.
}
\label{fig:ssns0}
\end{figure*}

Noticeably, the overlap between two cycles at the minimum has now reduced significantly
compared to the cases with lower diffusion (\Fig{fig:lowdifhr}).
Importantly, now we do not need to increase the observed flux distribution
by a large value 
to achieve sustained dynamo action; here $\Phi_0 = 2.4$.  
The amount of daily flux produced by the model with this value of $\Phi_0$ (\Fig{fig:hdref}(d)) 
is comparable to the observed BMR flux budget 
\citep[e.g.,][]{SH94,ZWL10,Li16}. 
Moreover, the average number of BMRs per cycle in this simulation is $3947$, 
which is very close to the observed group sunspot number ($3461$) obtained from 
the catalog of RGO and USAF/NOAA Sunspot averaged over the last 12 cycles (counting each spot only once).
Thus this is the first 3D solar dynamo model that is totally sustained by the observed distribution of 
tilted BMRs.

\begin{figure}
\centering
\includegraphics[width=1.0\columnwidth]{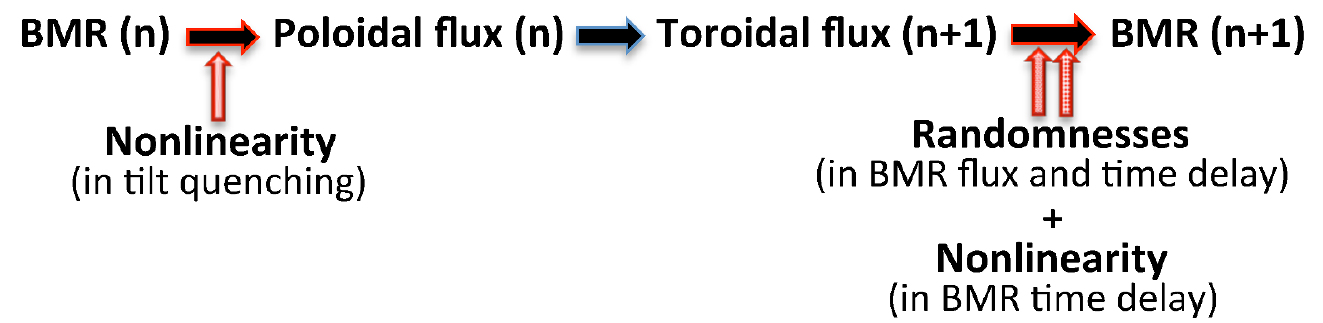}
\caption{
A schematic diagram of our BL dynamo model with nonlinearities and 
randomness involved into it (see text for discussion). Here ``n" refers to the cycle number.
}
\label{cartoon1}
\end{figure}

The magnitude of the magnetic pumping needed to sustain the dynamo (for example, $35$~\mps\ for $\Phi_0 = 1.5$ in Run~B6) 
is reasonable since it is still only a small fraction of the observed velocity amplitude of $1$--$2$ \kmps\
that characterizes solar surface convection \citep{NSA09}.
Notably, for the same value of diffusivity, 
our model uses less 
pumping to sustain 11~year dynamo cycle than the previous 2D \bl\ model of \citet{KC16}; see their Figure~16.
The possible reason could be the minor differences in the other parameters and the implementation of 
BL process (explicit BMR deposition vs. $\alpha$ coefficient).


We recall that the frequency of BMR emergences is governed by a
delay distribution of the type given in \Eq{eq:delaypdf}. This
produces a much more realistic variation of the surface BMR flux as
shown in \Fig{fig:hdref}(d).
Also the sunspot number (SSN) goes up and down with time in a similar fashion 
as the real sunspot cycle; see \Fig{fig:ssns0}.  We
remember that this is the actual SSN produced by the model and it
is not a proxy.  In all previous dynamo models \citep[e.g.,][]{CD00,JCC07,OK13,Pas14}, 
except the one of \citet{LC16}, a proxy of SSN is constructed based on the integrated
toroidal field near the base of the CZ.  Ours is the first 3D solar
dynamo model to explicitly produce a spontaneously-generated
distribution of BMRs that varies with the phase of the magnetic
cycle.

The asynchronous time delay of BMR emergence and the asynchronous flux distribution 
within two hemispheres
is sufficient to produce a considerable hemispheric 
asymmetry in the magnetic field and also in the BMR flux (\Fig{fig:hdref}).
The hemispheric asymmetry produced in this model is not much and gets corrected in one or two cycles. This
is expected because the diffusive coupling between two hemispheres at the equator
helps to reduce the hemispheric asymmetry. 
Furthermore, the stochastic process involved in the BMR emergence causes occasional
spikes at any phase of the solar cycle 
and sometimes causes double peaks in 
some cycles (e.g., around $605$ years and $645$ years in \Fig{fig:hdref}(d)).
Hemispheric asymmetry can also contribute to double peaks (see Fig.\ 15 below). 
Similar behavior is seen in many observed solar cycles \citep{McInt13}.

\begin{table}
\caption{Summary of the 
linear 
Pearson
correlation coefficients and the percentage significance level (s.l. = $(1-p)\times100\%$)
 between the polar flux of cycle n 
and the BMR flux (or the spot-producing bottom toroidal flux in the case of Run~A6) of the subsequent cycles.
}
\begin{center}
\begin{tabular}{llr}
\hline
Description of Run           & Correlation between            & Value (s.l. \%)\\
\hline
Advection-dominated,         & pol. flux (n) \& tor. flux (n)  &$0.18~(56.9)$ \\
and fluctuations in tilt     & pol. flux (n) \& tor. flux (n+1)&$0.64~(99.9)$ \\
(Run~A6).                    & pol. flux (n) \& tor. flux (n+2)&$-0.23~(78.6)$ \\
                             & pol. flux (n) \& tor. flux (n+3)&$-0.19~(49.8)$ \\
\hline
Diffusion-dominated,         & pol. flux (n) \& BMR flux (n)  & $0.79~(99.9)$ \\
and no fluctuations in       & pol. flux (n) \& BMR flux (n+1)& $0.88~(99.9)$ \\
tilt (Run~B9).               & pol. flux (n) \& BMR flux (n+2)& $0.77~(99.9)$ \\
                             & pol. flux (n) \& BMR flux (n+3)& $0.66~(99.9)$ \\
\hline
Same as Run~B9 but           & pol. flux (n) \& BMR flux (n)  & $0.87~(99.9)$ \\
with fluctuations in         & pol. flux (n) \& BMR flux (n+1)& $0.97~(99.9)$ \\
 tilt (Run~B10).                   & pol. flux (n) \& BMR flux (n+2)& $0.84~(99.9)$ \\
                             & pol. flux (n) \& BMR flux (n+3)& $0.75~(99.9)$ \\
\hline
Same as Run~B10 but          & pol. flux (n) \& BMR flux (n)  & $0.37~(99.8)$ \\
four times weaker $\Bsat$    & pol. flux (n) \& BMR flux (n+1)& $0.86~(99.9)$ \\
(Run~B12).                   & pol. flux (n) \& BMR flux (n+2)& $0.25~(95.5)$ \\
                             & pol. flux (n) \& BMR flux (n+3)& $0.00~(21.0)$ \\
\hline
Same as Run~B10 but          & pol. flux (n) \& BMR flux (n)  & $0.71~(99.9)$ \\
diffusivity in the tachoc-   & pol. flux (n) \& BMR flux (n+1)& $0.91~(99.9)$ \\
line is same as the value    & pol. flux (n) \& BMR flux (n+2)& $0.58~(99.7)$ \\
in the CZ (Run~B13).         & pol. flux (n) \& BMR flux n+3& $0.41~(93.5)$ \\

\hline
\end{tabular}
\end{center}

\label{table2}
\end{table}

Despite the tilt angle quenching, the model produces
an observable variation in the amplitude of the cycle. This is particularly seen in the daily BMR flux of \Fig{fig:hdref}(d)
and in the monthly SSN (\Fig{fig:ssns0}).
The amount of variation in the peak monthly SSN is $\approx 14\%$.
We recall that in this model there is no randomness in the tilt angle around Joy's law. 
Thus we need to consider what other factors give rise to the cycle variability.

We address this issue with the schematic diagram shown in \Fig{cartoon1}. 
In the BL process, decay and dispersal of tilted
BMRs on the solar surface produce poloidal field
at the end of the cycle. Thus we expect the polar flux of a cycle
to depend on the amount of flux that has emerged in BMRs during that cycle 
and we expect these two quantities to be highly correlated.
However, we get a linear correlation coefficient of less than $1$; 
see the 2nd row in \Tab{table2} for all correlations.
The reason behind the reduction of the correlation is the nonlinearity 
in the tilt angle which reduces the tilt when the BMR field exceeds $\Bsat$.
This nonlinearity is shown by the first vertical arrow in \Fig{cartoon1}.
The variation in the mean BMR latitudes has also some effect
in the process: BMR~(n) $\rightarrow$ Poloidal flux~(n), 
although in this simulation there is not much variation of it
and we ignore it in the discussion.

The poloidal field produced on the solar surface is transported to the deep CZ where
differential rotation produces a toroidal field for the next cycle. Thus the process:
Poloidal flux~(n) $\rightarrow$ Toroidal flux~(n+1) is fully deterministic.
The next process, Toroidal flux~(n+1) $\rightarrow$ BMR~(n+1),
however, is not fully deterministic because both the
BMR time delay and BMR flux are taken randomly
from their distributions. These sources of randomness are indicated by the second vertical arrow in \Fig{cartoon1}. 
However, they
largely average out over many BMRs; otherwise, we would not get a
strong correlation between the polar flux~(n) and the BMR flux~(n+1) as listed in \Tab{table2}.  
This is in agreement with the
correlation obtained from the observed polar field data \citep{CCJ07}
and from different proxies of the polar field \citep{Muno13,Priy14}.
In fact, this correlation is a popular basis for the solar cycle
prediction \citep{Sch78}.

We must remember that although the polar flux~(n) and thus the toroidal flux~(n+1) 
is positively correlated with the BMR flux~(n+1), the process may not be linear. 
In our model, the BMR delay distribution involves a nonlinearity---it produces
more BMRs when the toroidal flux at the base of the CZ is stronger.
This nonlinearity is identified by the third arrow in \Fig{cartoon1}.

From the above analysis, we realize that the causes of the magnetic
cycle variation in this model are the nonlinearities in tilt angle
and in the delay distribution, and the randomness in the BMR emergence
process. As discussed above, the randomness in the BMR emergence
has a minor contribution to the cycle variation, although it is
difficult to separate out the contributions of each component.

\subsubsection{Solution with observed tilt angle fluctuations}
\begin{figure}
\centering
\includegraphics[width=1.00\columnwidth]{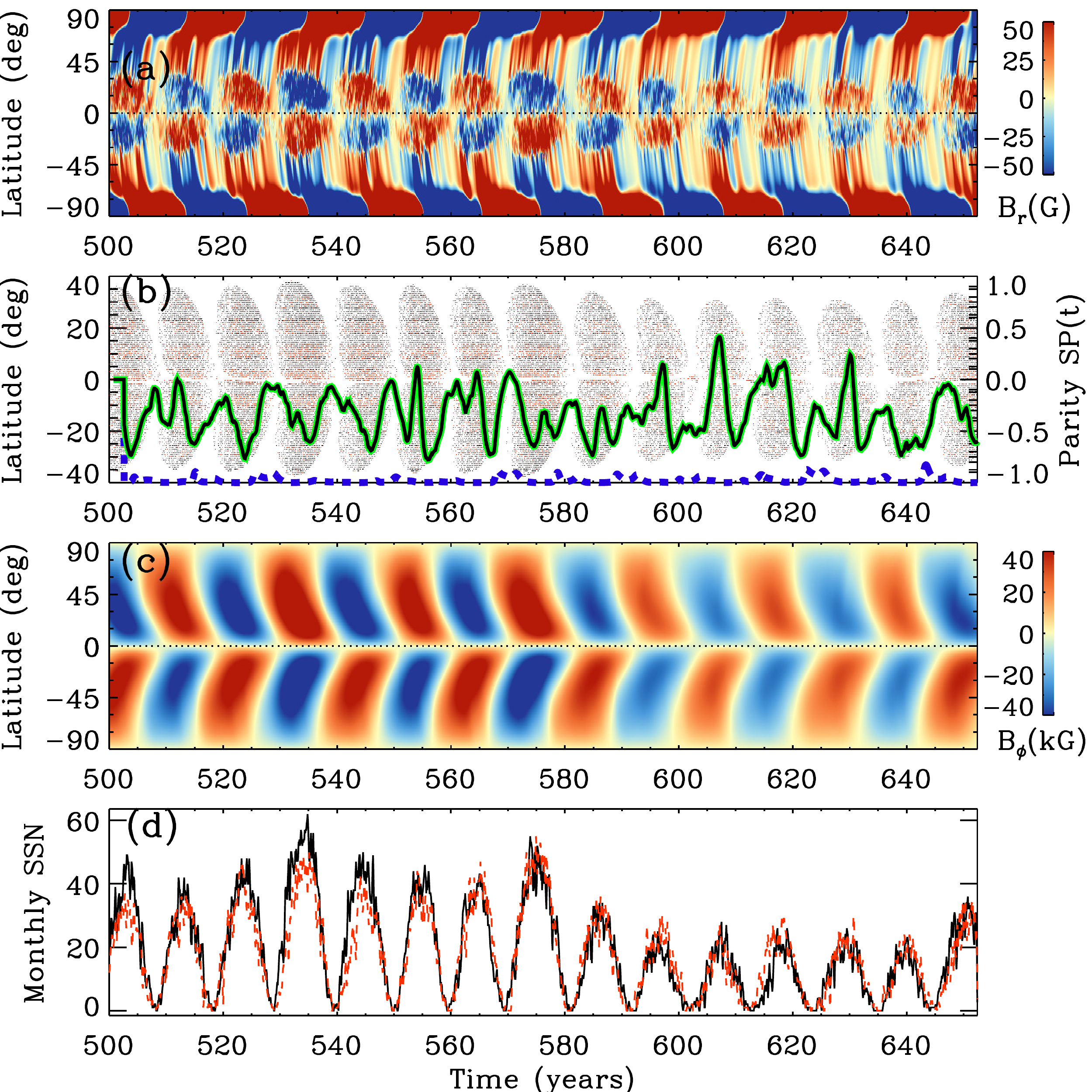}
\caption{
Results from Run B10: temporal variations of (a) $\brac{B_r(R,\theta,\phi)}_\phi$, 
(b) latitudes of BMRs, (c) $\brac{B_\phi(0.72R,\theta,\phi)}_\phi$, and (d) the monthly smoothed SSNs; black/red: north/south. 
In (b), red points show the wrongly tilted BMRs; green/solid and blue/dashed lines show parities, SP(t) computed over the four years of 
surface $B_r$ and the bottom $B_\phi$, respectively.
}
\label{fig:hdflc}
\end{figure}

\begin{figure*}
\centering
\includegraphics[width=2.10\columnwidth]{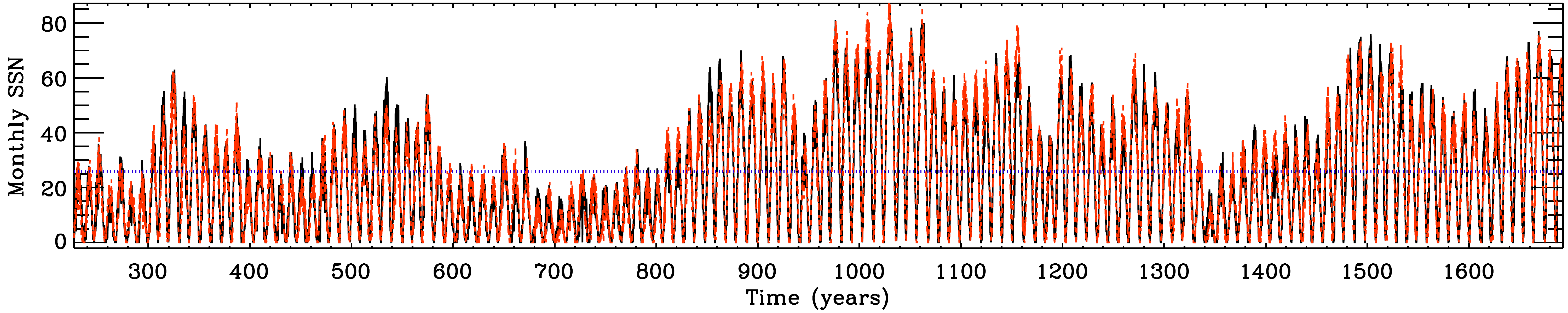}
\caption{The monthly BMR number (smoothed over three months) with time. This is
obtained from Run~B10 which is displayed in \Fig{fig:hdflc} but after running it for a longer time.
The dotted line shows the mean of the observed peak SSNs for last $13$ cycles.
}
\label{fig:hdrefssn}
\end{figure*}

In the above model, we now include variation in the tilt angle as
guided by the observation, i.e., a Gaussian fluctuation with
$\sigma_\delta=15^\circ$ around Joy's law (\Eq{eq:joys}). 
Run~B10 in \Tab{table1} refers this case. A few
cycles from this stochastically driven dynamo simulation are presented
in \Fig{fig:hdflc}, while the sunspot time series from the full
simulation is shown in \Fig{fig:hdrefssn}.  Comparing \Fig{fig:hdflc}
with \Fig{fig:hdref}, we notice a greater variation in the
magnetic field.  Particularly, in \Fig{fig:hdflc}(a) we observe
frequently mixed polarity field as a consequence of the wrong tilt.  The cycle-to-cycle
variation of the amplitudes of the mean polar flux: $\sqrt{
\frac{1}{N}\sum_{i=1}^N ({\overline{B}_r}_i - \overline{B}_r^{\rm
avg}})^2 /\overline{B}_r^{\rm avg} \times 100 \% \approx 35\%$
(where $N = 93$).  
This value is in agreement with \citet{JCS14} who found about $30\%$ variation for the cycle $17$ 
in the axial dipole moment and the polar field 
compared to the value without tilt scatter. 

The
strength of the magnetic field and the number of BMRs per cycle have increased in
this simulation with respect to the simulation without tilt fluctuations
(Run~B9); see \Tab{table1}. The reason for this will be explored later.  
The amount of variation in the peak
SSN: $\sqrt{ \frac{1}{N}\sum_{i=1}^N (PSN_i - \overline{PSN})^2
}/\overline{PSN} \times 100 \% \approx 41\%$, while in the observed
data (http://www.sidc.be/silso/datafiles) for $1749$--$2017$,
it is $32\%$.  As the variation of the SSN in this model
is much larger than that obtained from the model without tilt 
fluctuations, we can certainly conclude that the fluctuations in
the flux emergence process and the nonlinearity in the \bl\ process
have a relatively minor effect on the variation of the magnetic 
cycle relative to the tilt angle scatter.  Furthermore, this suggests 
that the observed tilt angle scatter in the Sun may be sufficient to 
account for the observed solar cycle variability.

\begin{figure}
\centering
\includegraphics[width=1.0\columnwidth]{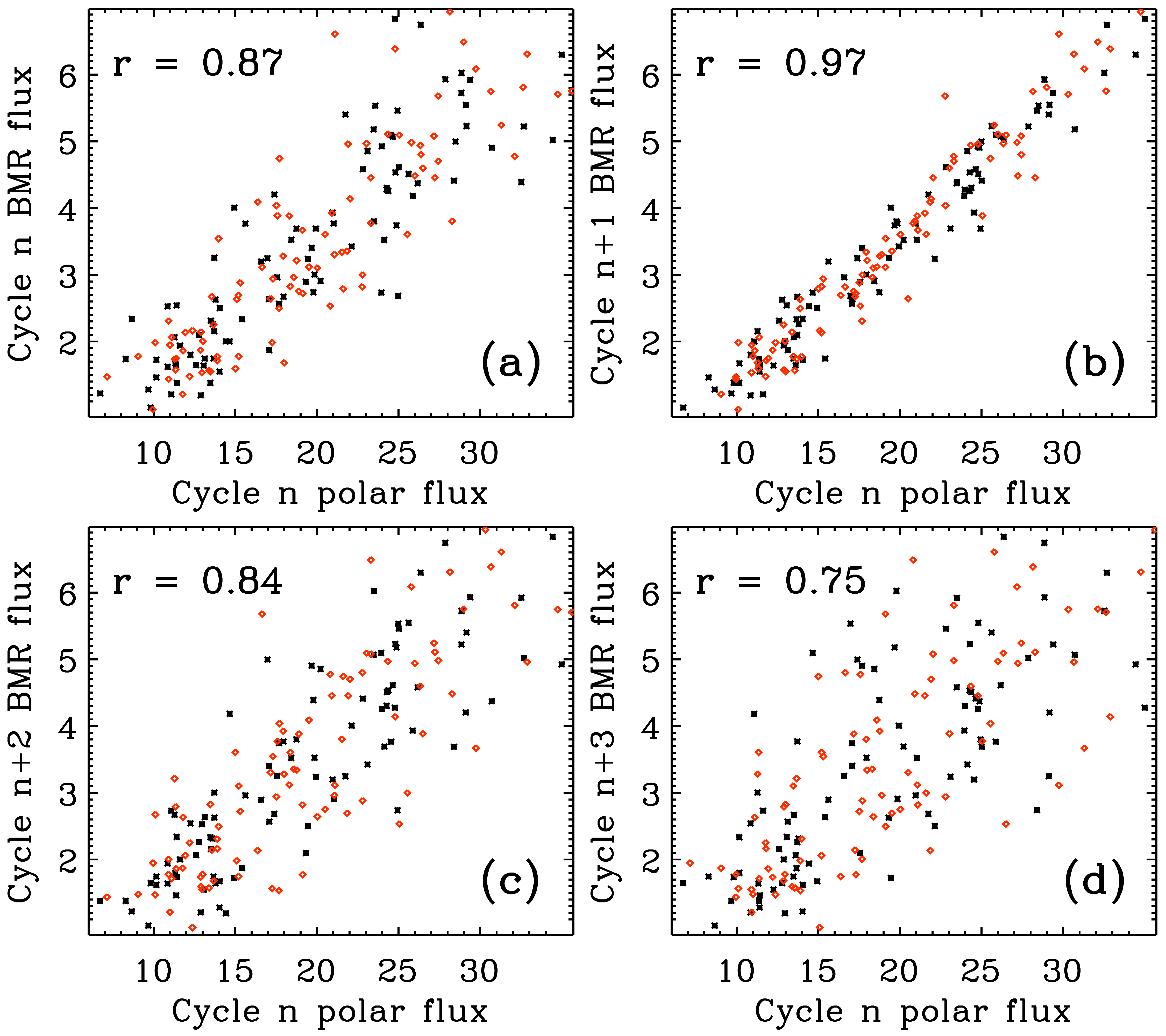}
\caption{
Scatter plots between the polar flux density (G)
of cycle n and the daily BMR flux (in unit of $10^{22}$~Mx) of (a) cycle
n, (b) cycle n+1, (c) cycle n+2, and (d) cycle n+3 from Run~B10.
Significance levels of all correlations are above $99.9$\%.
Two different symbols correspond to two different hemispheres.
}
\label{fig:hdcorr}
\end{figure}

The basic dynamo loop shown in \Fig{cartoon1} still applies for this model but with 
the inclusion of a randomness due to tilt scatter in the process: BMR (n) $\rightarrow$ Polar flux (n).
Interestingly, we still find a fairly good correlation (r = 0.87) between 
the
BMR flux (n) and the polar flux (n); see \Fig{fig:hdcorr}(a). 
Using the polar faculae as a proxy
for the polar flux, \citet{Muno13} find a little correlation between
the polar flux and the SSN of the same cycle.
If their result is true, then it suggests that in the \bl\ process of our model,
the nonlinearity and randomness are weaker than in the real Sun.

As obtained from the previous model without tilt fluctuations,
a strong correlation between the polar flux (n) and the
BMR flux (n+1) is expected as shown in \Fig{fig:hdcorr}(b).
We remember that this
correlation is very robust and a similar correlation is obtained if we consider the peak SSN instead of the peak BMR flux.
Moreover, a similar correlation is also obtained from the previous advection-dominated model; see \Tab{table2}.
This is consistent with the idea 
that a reliable prediction of the future solar cycle is possible
using the observed polar field of the previous solar minimum
\citep{Sch78,CCJ07,JCC07}.

As in the process: BMR flux (n) $\rightarrow$ Polar flux~(n), the correlation is not completely broken,
the polar flux still has a correlation with the BMR flux~(n+2).
This is shown in \Fig{fig:hdcorr}(c).
This correlation gets weakened in each transformation: poloidal flux~(n) $\rightarrow$ BMR~(n+1) $\rightarrow$ poloidal flux~(n+1).
Hence, we get a much weaker correlation between the polar flux~(n) and BMR flux~(n+3).

\citet{JCC07} and \citet{YNM08} concluded that the memory of the polar flux is determined by the relative importance of diffusive and advective flux transport. In the diffusion-dominated model, they find one cycle memory between the polar flux and the toroidal flux,
while in the advection-dominated it is three cycles. However, we find that the memory of the polar flux
is not primarily related to the flux transport process, rather it is a fundamental consequence of any cyclic \bl\ process.
As explained through \Figs{cartoon1}{fig:hdcorr}, if the correlation between the BMR flux and the polar flux of the same cycle is not completely broken, then this correlation has to propagate for many cycles.
This has happened in Figure 12 of \citet{YNM08} what they identify as the advection-dominated model.
However in Figure 11 of \citet{YNM08} the same cycle correlation has been broken and they called this as the diffusion-dominated regime.
The broken correlation in their case is due to diffusion, while in our case is due to both the nonlinearity in the \bl\ process and the diffusion.
This is confirmed by repeating the same simulation as shown in \Fig{fig:hdcorr} but by reducing the
$\Bsat$ of the tilt angle quenching in \Eq{eq:joys} by four times (Run~B12). The correlations between different cycles are listed in \Tab{table2}.
As we can see from \Eq{eq:joys} that when we keep everything else same in the model but reduce $\Bsat$, the nonlinearity in the model effectively increased. This
nonlinearity in the tilt angle acts to break the linear dependence between the polar flux and the BMR flux of the same cycle.

\begin{figure*}
\centering
\includegraphics[width=2.05\columnwidth]{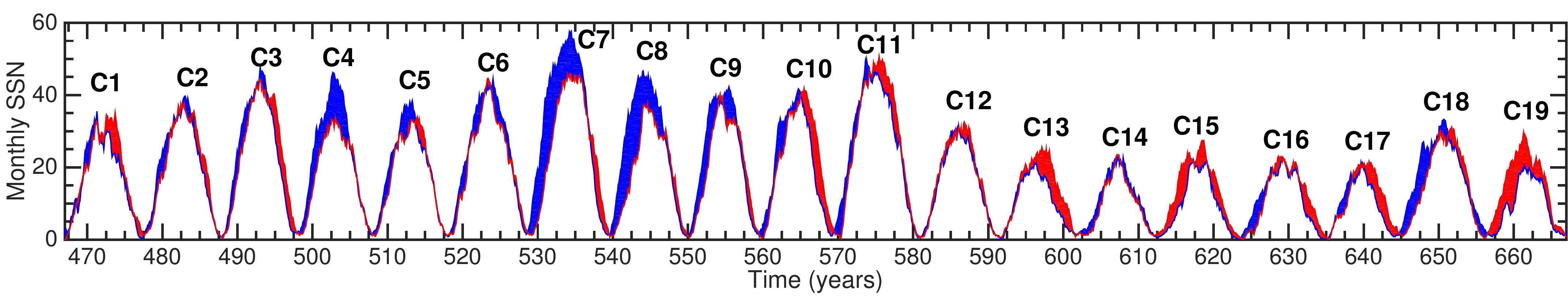}
\caption{
A portion of the smoothed SSN time series shown in \Fig{fig:hdrefssn} with the red and blue show the northern and 
southern hemisphere numbers, respectively. The shading area represents the excess of the BMRs 
between two hemispheres. To facilitate the discussion, we have labeled the cycles.
}
\label{fig:zoomedBMR}
\end{figure*}

One may think that a much weaker diffusivity in the tachocline has made our model more like the advection-dominated model 
and might be the cause of many cycles correlations in \Fig{fig:hdcorr}.
To check this we have performed another simulation by increasing the tachocline diffusivity
to $1.5\times10^{12}$~\cmss, i.e., $\etat$ in the tachocline is now same as in the CZ (Run~B13).
No other changes are made in this simulations with respect to Run~B10.
Again in this simulation, we find similar values for correlations as listed in the last row of \Tab{table2}.
Stronger diffusion in the tachocline tries to reduce the correlation in each cycle but never diminishes it
to one cycle as we expect in the diffusion-dominated region.

We also mention that \cite{KN12} find a reduction of the memory in both advection- and diffusion-dominated
dynamos to one cycle by the inclusion of a downward pumping. Actually, the pumping increases the strength of the magnetic flux
and thus the nonlinearity, which reduces the memory to one cycle in \cite{KN12}.

Thus to summarize the whole idea; in the \bl\ dynamo, as long as there is an efficient mechanism to transport
the surface poloidal flux to the deep CZ, the polar flux and the BMR flux are cyclically coupled (\Fig{cartoon1}).
If the nonlinearity in the \bl\ process or the relative diffusive transport is sufficiently strong, then the memory of the polar flux will be limited to the next one cycle only,
otherwise, it will be propagated to multiple cycles.

\begin{figure}
\centering
\includegraphics[width=1.0\columnwidth]{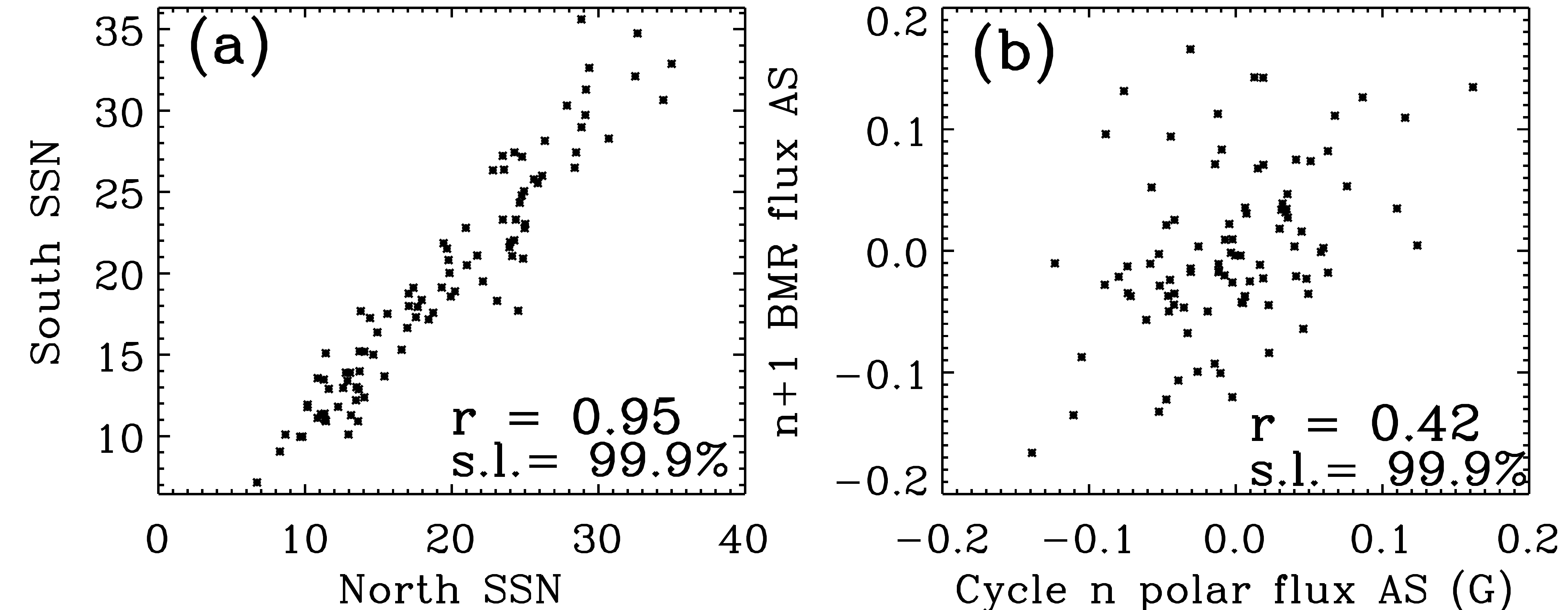}
\caption{Scatter plots between (a)
peaks of the northern SSN and the southern SSN,  
(b) the normalized hemispheric asymmetry of polar flux of cycle n 
and the asymmetry of BMR flux of cycle n+1.
}
\label{fig:hdasym}
\end{figure}

\begin{figure}
\centering
\includegraphics[width=1.0\columnwidth]{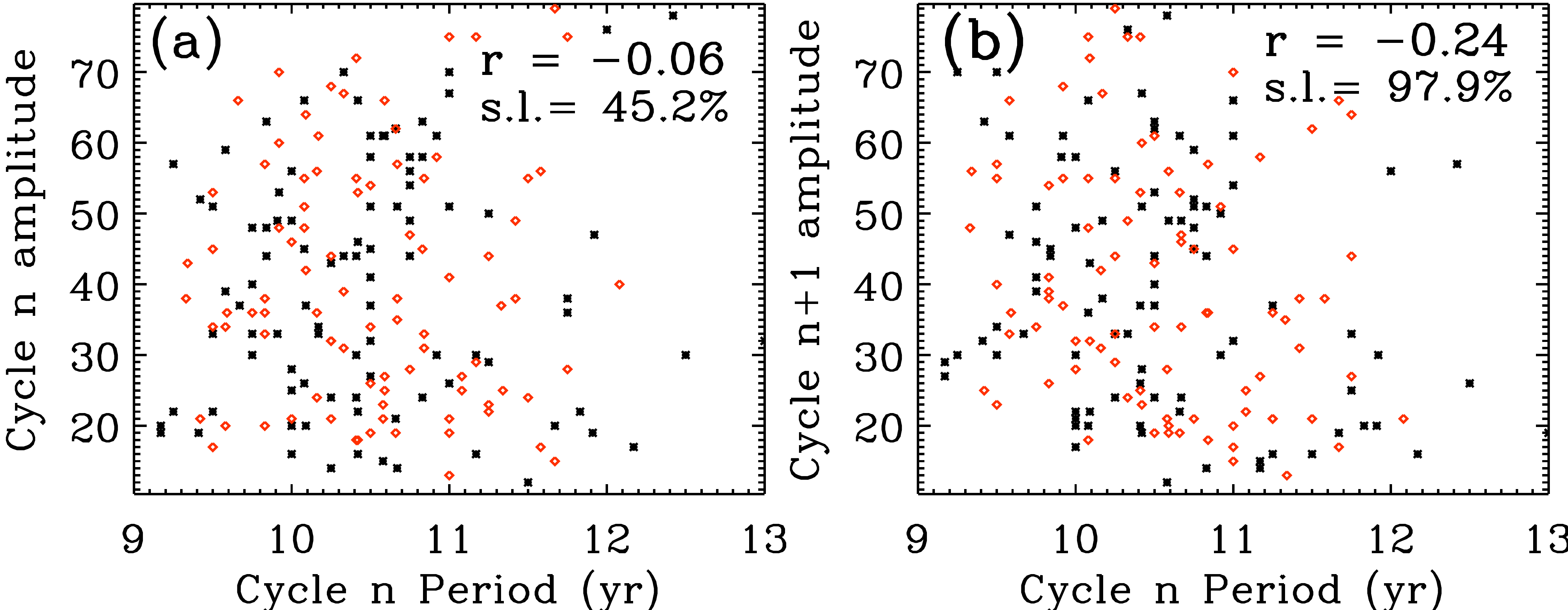}
\caption{Scatter plots between periods and amplitudes of (a) the same cycles, and (b) the next cycles.
Two different symbols correspond to two different hemispheres.}
\label{fig:pervsamp}
\end{figure}

Going back to the SSN plot in \Fig{fig:hdrefssn}, we observe some hemispheric asymmetry.
In \Fig{fig:zoomedBMR}, we highlight it for some cycles.
In this figure, we clearly observe the temporal lag and the excess of BMRs between two hemispheres.
We notice that first three cycles in this figure are more or less symmetric. 
Then in cycle C4, the southern hemisphere got more spots, although the temporal symmetry is still retained. 
In the next cycle, the excess of spot in southern hemisphere
has now reduced and eventually in C6, it has diminished completely. Again in C7, a new asymmetry is introduced. 
But now the southern hemisphere has more spots and this hemisphere is leading over the other in the rising phase. This is continued for the next two cycles. Then for C10--C13, the northern hemisphere has got little more spots, particularly during the decaying phase. C14 is very symmetric, while for C15, the northern hemisphere is little longer than the other. Finally, for C18, the southern hemisphere is leading in the rising phase, while for C19, it is opposite.

Certainly, we cannot make a one-to-one comparison of our sunspot cycles with the observed ones as we do not model the exact observed cycles.
However, on comparing our sunspot cycles in \Fig{fig:zoomedBMR} with observed cycles in Figure~10 of \citet{McInt13},
readers can convince yourself that very similar features of the solar cycle are reproduced in our model.

We have seen in \Fig{fig:zoomedBMR} that like the Sun, our model always
tends to correct any (hemispheric or temporal) asymmetry produced in a cycle
and we do not observe extended asymmetry.
Hence we obtain a strong correlation between
the amplitudes of the north and the south sunspot cycles as
shown in \Fig{fig:hdasym}(a).  The polar flux asymmetry obtained
in this diffusion-dominated model is comparable to the value obtained
from the previous advection-dominated model; compare the horizontal axes
of \Fig{fig:hdasym}(b) and \Fig{fig:lowdifasy}(a). However, the
correlation between the polar flux asymmetry with the SSN asymmetry of the next cycle (\Fig{fig:hdasym}(b)) is much
less than that found in the previous advection-dominated model.
This is expected because in the diffusion-dominated model,
fields are largely coupled across the equator
and much of the memory of the polar flux asymmetry does not preserve in the toroidal flux.
Moreover, due to asynchronous BMR emergence process, a new asymmetry is introduced 
(which is not related to the polar flux asymmetry).

In \Fig{fig:pervsamp}(a) we show the scatter plot between
the amplitudes and the periods.  While in observations \citep{CD00}, there is a
little anti-correlation, in our model we find almost no correlation.
Interestingly, from the horizontal axis of this figure, we notice that
the cycle period has considerable variation around its mean of $10.5$ years.
In the flux transport dynamo paradigm, we believe that the cycle period is
largely determined by the speed of the meridional flow \citep{DC99}, which is kept constant here.
Thus the variation in our period is caused by the fluctuations and nonlinearities in the BMR emergence.
Let us discuss how this is happening.
When the polar field of a cycle becomes stronger due to the tilt fluctuations,
spots in the next cycle take a longer time to reverse the previous cycle flux. 
This effect acts to make the cycle longer. However, there is another counter effect.
Stronger polar flux makes the toroidal flux stronger which makes more frequent 
BMR emergence. This effect acts to reverse 
the polar flux quickly and makes the cycle period shorter, though it is inhibited by the tilt angle quenching.
The competition between these two effects causes variation in the period.

Finally, we find a little anti-correlation between 
amplitudes and periods of previous cycles as shown in
\Fig{fig:pervsamp}(b). In observations \citep[see e.g.,
Figure 4 of][]{Haz15} this correlation is $-0.67$, 
which is much larger than our value.

\begin{figure*}
\centering
\includegraphics[width=2.10\columnwidth]{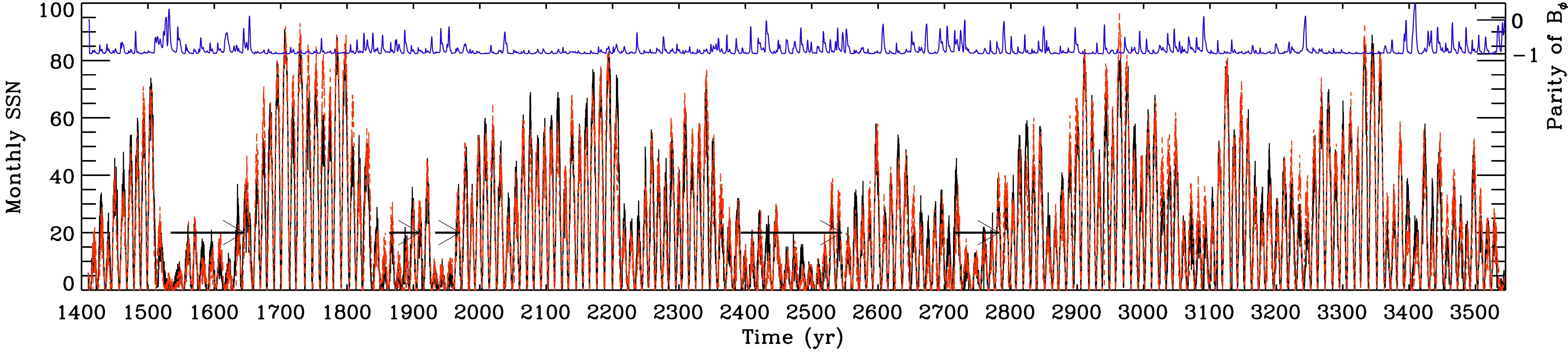}
\caption{SSN time series (black/red: north/south) from Run~B11 i.e., with $\sigma_\delta=30^\circ$.
The top blue line shows the evolution
of the smoothed (over $11$~years) parity SP$_\phi$(t) of the bottom toroidal field.
Arrows show the extent of grand minima based on our definition.
}
\label{fig:30sigma1}
\end{figure*}

\subsubsection{Grand minima and Maxima}
In \Fig{fig:hdrefssn}, we have seen that a random component following
a Gaussian distribution with $\sigma_\delta=15^\circ$
around Joy's law occasionally produces very
weak and strong cycles, and a few Dalton-like extended
period of weaker activity (e.g., around $700$ years in \Fig{fig:hdrefssn}).
Yet, the dynamo never becomes so weak to produce any Maunder-like grand minimum. 
However, we must remember that for all BMRs we have considered
the same level of tilt fluctuations, while in
observations, there are indications that weaker BMRs have bigger
scatter in their tilts \citep{SK12,JCS14,LCC15}. 
Moreover,
the tilt variation that we have
implemented in above simulations, is extracted from the variation within
a solar cycle data (for example, cycle $23$ in the analysis of
\citet{SK12} and cycle $21$ in the analysis of \citet{LCC15}).  In observations \citep[e.g.,][]{Das10,Wang15,Arlt16},
we find the tilt to have cycle-to-cycle
variation in
addition to variations within a cycle. Motivated
by these observational results, we double the tilt fluctuations, i.e., we now take $\sigma_\delta = 30^\circ$
instead of $15^\circ$.  
This simulation is labeled as Run~B11 and the sunspot time series from this
simulation is shown in \Fig{fig:30sigma1}.  Interestingly, again
the dynamo does not shut off and the cycle is still maintained even
at this large tilt fluctuations.  We find several
episodes when the magnetic field and the cycle become much weaker,
for example, around $1600$, $1900$, and $2500$ years. These events can
be considered as Maunder-like grand minima.

To compute the number of grand minima and the time spent in those events,
we follow the same procedure as applied in \citet{USK07}. We first bin the data in $10$ years interval and then
filter the data using the Gleisberg's low-pass filter 1-2-2-2-1. 
We consider a grand minima when SSN goes below $50\%$ of the mean at least for two consecutive decades.
Applying this procedure in the previous data set of Run~B10 with $\sigma_\delta=15^\circ$, 
we now get two grand minima (around times $1950$~year and $2600$ year).
This simulation spent $9.3$\% of its time in these grand minima phase which
is much less than the value of $17$\% obtained in $^{14}$C data of \citet{USK07}.
This simulation also produces two grand maxima with time spent in these phases
is $7.6$\% which is again less than the value of $9$\% obtained in $^{14}$C data.
 
Ironically, the simulation of $\sigma_\delta = 30^\circ$ produces
$26$ grand minima in $11400$~years of the simulation run. Out of these $26$ grand minima,
five are shown in by arrows in \Fig{fig:30sigma1}. 
Our number of grand minima is very close 
to the value $27$, obtained in the last $11400$ years of $^{14}$C data \citep{USK07}.
The time spent in the grand minima is 18\% which is again very close
to the record from $^{14}$C data.
We are carrying out a detailed analysis of the grand minima, particularly how our model
recovers from grand minima phase, owing to
a few BMRs. These will be presented in a forthcoming publication.

Our model also produces occasional periods of stronger activity
resembling the solar grand maxima. In this simulation we obtained
$17$ grand maxima with time spent in these phases is $9.6$\%.
Again these values are close to the ones obtained in $^{14}$C data.
A detailed study of grand maxima
will also be presented in the forthcoming publication. 

On comparing Runs~B9--B11 in \Tab{table1}, we notice that
$\overline{\rm SP}_\phi$
increases with the increase of
the tilt angle scatter ($\sigma_\delta$), i.e., going towards the
quadrupolar parity from the dipolar one. Also the
deviation from the dipolar mode, as seen by the value of ${\sigma_{\rm SP}}_\phi$, increases with
the scatter. It is not difficult to understand the reason.  Due to
scatter in the tilt, when a BMR gets wrong tilt
in one hemisphere, it produces a quadrupolar field instead of
a dipolar field. The occurrence of this event increases with the increase
of tilt scatter and thus the parity tends to go to the quadrupolar parity.
During grand minima when there are less number of BMRs, the
effect of tilt fluctuations is more pronounced and the deviation of the dipolar parity
is significant, as seen in \Fig{fig:30sigma1}.

\begin{figure*}
\centering
\includegraphics[width=2.10\columnwidth]{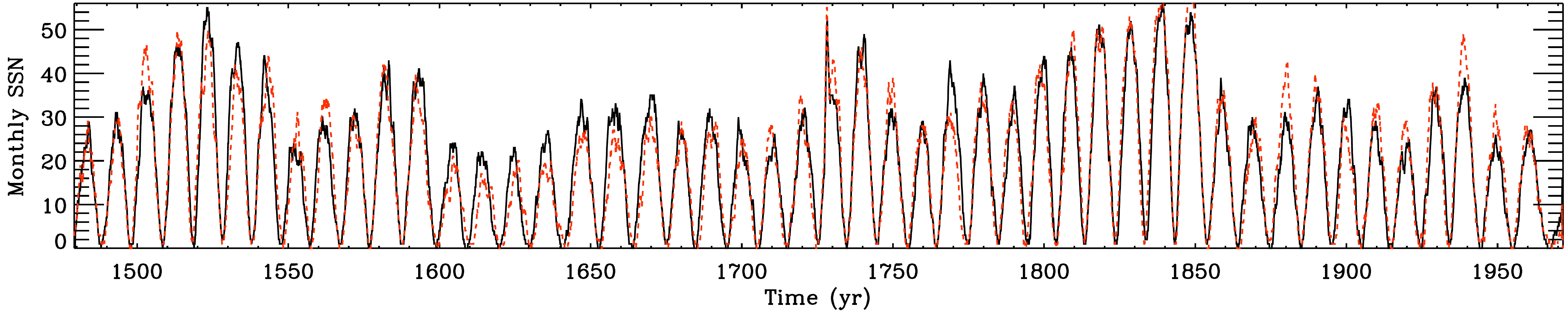}
\caption{Same as \Fig{fig:hdrefssn} (Run~B10), but this is obtained from Run~C1 in which the quenching is operating 
in BMR flux and not in the tilt.
}
\label{fig:case1}
\end{figure*}

\begin{figure*}
\centering
\includegraphics[width=2.10\columnwidth]{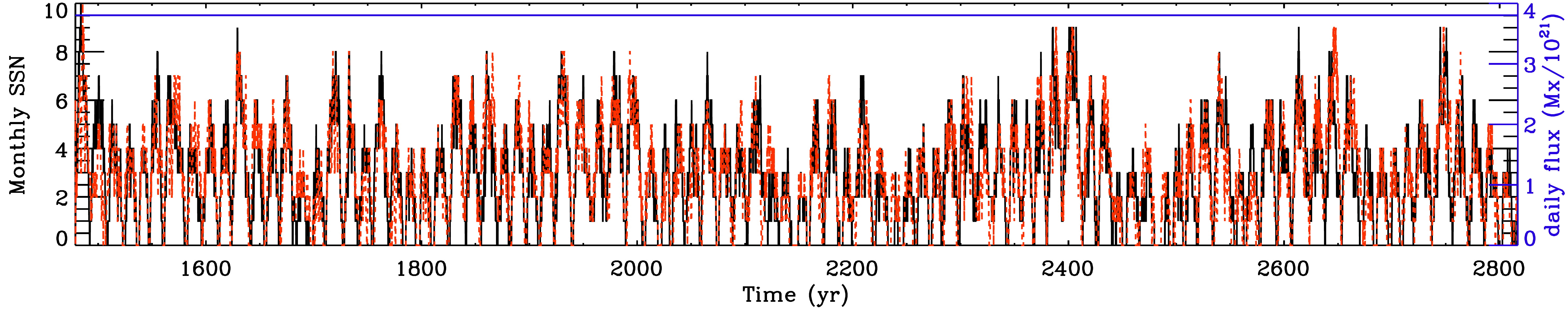}
\caption{Same as \Fig{fig:hdrefssn} (Run~B10), but obtained
from Run~D1 in which the magnetic field-dependent nonlinearity on
the BMR delay is different; see \S\ref{sec:diffnonlin} for details.
The horizontal line shows the mean daily BMR flux (in the unit of $10^{21}$~Mx) 
in this model.}
\label{fig:case2} \end{figure*}

\subsection{Sensitivity of solutions with nonlinearities}
\label{sec:diffnonlin}

To explore the sensitivity of the solar cycle variation with
nonlinearities in the model, we consider Run~B10 and we perform following two new simulations.
First, instead of taking the nonlinear quenching factor:
$1/[1+(\hat{B}/\Bsat)^2]$ in the tilt angle (which is the case in
all previous simulations), we take it in the BMR flux. The Run~C1
in \Tab{table1} represents this case.  Second, we keep the tilt
quenching as before but change the magnetic field dependent factor
$1 / [1 + ({B_b^N}/{B_\tau})^2]$ in \Eq{eqtau} for $\taup$ and
$\taus$ to $1 / (1 + {B_b^N}/{B_\tau})$. The Run~D1 represents
this case.

The result for Run~C1 is shown in \Fig{fig:case1}. As listed
in \Tab{table1}, the period and the number of BMRs per cycle are
smaller in this simulation, although the morphology of the field
(not shown) looks very similar to the previous simulation (Run~B10).
However, the variation of the peak SSN is $29\%$ which is somewhat 
smaller than in Run~B10 ($41\%$).  It is surprising that on putting the same quenching
factor from the tilt to the flux, the model produces a different
amount of variation in the solar cycle. The reason is that when the
quenching is operating in the flux, the dynamo becomes more stable
than when it is operating in the tilt. 
To clarify this point, we
first estimate the magnetic field generated from only two symmetric BMR
pairs deposited at $\pm5^\circ$ latitudes at the beginning of a simulation. 
Tilts of these pairs are given by Joy’s law and no seed magnetic field is given in
this simulation. (This study is very similar to the one presented in Section 4.2 of \citet{HCM17}.)
Then we perform two more simulations.
In one, we reduce the flux of pairs by $50\%$ and in another, 
we keep the flux same but reduce the tilt by the same amount.
After running these simulations for about $5$~years, we find that the
high-latitude radial flux in the former case has reduced by about
$64\%$, while in the latter case it is reduced by only $50\%$.  Thus
when the magnetic field tends to grow, it is easy for the dynamo
to stabilize it by reducing the BMR flux than reducing the same amount
of tilt.  This conclusion becomes even stronger by comparing values
of $\tilde{B}_{tor}$, $\tilde{B}_{r}$, and the mean BMR number per
cycle for Runs~B10 and C1 in \Tab{table1}.  We notice that all these
values are smaller in Run~C1, confirming that the flux quenching
did not allow the field to grow much.

Finally, we consider the Run~D1 which produces a decaying solution unless we increase the
flux distribution by a small amount (see \Tab{table1}). The solution,
in this case, shows a considerably different behavior.  The overlap
between cycles at the minimum has increased and we do not observe
very distinct cycles; see \Fig{fig:case2}.  Moreover, the mean period
becomes longer ($14$~years instead of $10.5$~years as in Run~B10) and
cycles are very irregular.  This is expected because on decreasing
the strength of the nonlinearity in $\taup$ and $\taus$, the rate
of spot production decreases and thus the polarity reversal becomes
slower. The most significant feature in this simulation is that the
dynamo is still operating with only a few BMRs although the BMRs
are little bigger (due to their larger flux). Thus
$\tilde{B}_{tor}$, $\tilde{B}_{r}$, and the mean value of the daily
BMR flux (the horizontal line in
\Fig{fig:case2}) are also less compared to the previous Run~B10.
From this simulation, we can conjecture that this scenario might be
applicable to other stars (probably the slowly rotating stars) which
produce fewer BMRs, and irregular and overlapped cycles.  Another
point to note is that the variation of the peak SSN in this
simulation is less than in Run~B10. We expect the variation to be
larger due to the smaller number of BMRs but because of having less 
sensitive spot production rate with the magnetic field, the sunspot 
variation is reduced. 
From this simulation
we learn that with the decrease of the sensitivity of $\taup$
and $\taus$ with the magnetic field, the cycle-to-cycle variation
in the peak SSN decreases slowly.

\section{Summary and Conclusions}

Using observed properties of the BMRs in our previous 3D \bl\ solar dynamo model (MD14 and MT16), 
we have studied the behavior of the dynamo action and the causes of the variability of the magnetic cycle in this model.

We take the flux of BMRs from an observed distribution \citep{Mu15} and then we couple the net surface flux budget of BMRs with the toroidal field at the base of the CZ. We do this in two ways. First, we scale the observed flux distribution based on the toroidal flux at the base of the CZ (\Sec{sec:fixedTD}). In this case, the delay distribution of BMR emergence is held fixed. Although the net BMR flux produced by this model has some variation with the magnetic cycle, the SSN does not show appreciable variation due to a considerable overlap between two magnetic cycles at the minima (\Fig{fig:LDzoomed}).

In the second approach, we keep the observe flux distribution unchanged but vary the BMR emergence rate based on the toroidal flux at the base of the CZ (\Sec{sec:BdepTD}). 
Thus we get more BMRs at the solar maximum when the toroidal field is strong. As a result, we attain cyclic variations in the BMR flux and in the BMR number, in the same manner as we observe in the Sun.
Thus for the first time in our model, we obtain a sunspot cycle that can be compared directly with observations, as opposed to using a proxy for this (e.g., \Fig{fig:ssns0}). Our main results are itemized below.

\begin{description}
\item[$\bullet$ ]
The overall dynamo growth is limited by a nonlinearity in the tilt angle. This is the only nonlinearity in the model when the time delay distribution is fixed.

\item[$\bullet$]Reduction of the tilt angle by only a few degrees is sufficient to limit the dynamo growth (\Fig{fig:hdtilt}). 
Thus, potential signatures of tilt quenching in solar observations may be subtle.

\item[$\bullet$ ]When the BMR delay distribution is nonlinearly coupled with the toroidal flux,
this nonlinearity acts in counter to the tilt angle nonlinearity. In contrast to tilt nonlinearity,
the delay nonlinearity acts to make the poloidal field strong by producing more BMRs when the toroidal field becomes strong.
Thus the variation of the magnetic cycle in our model is controlled by the competition between these two nonlinearities.

\item[$\bullet$ ]
These two nonlinearities, along with the randomness in the BMR emergence, are capable of producing
a substantial variation in the magnetic cycle, as reflected by the SSN (\Fig{fig:hdref}). 
A noticeable hemispheric asymmetry is also observed in this model.

\item[$\bullet$ ]
The variability of the magnetic field is more when the BMR delay distribution is dependent on the magnetic field; compare $\overline{B_r}$ from A series of simulations with other simulations in \Tab{table1}.

\item[$\bullet$ ]
When a scatter in the BMR tilt around Joy's law is included, the model produces much larger variation
in the magnetic cycle.
The cycle variability in our simulations for $\sigma = 15^\circ$, ranges from $19$--$59$$\%$, 
depending on the flux transport (diffusion and pumping), 
and on the nonlinearities in the BMR emergence rate and tilt angles; see \Tab{table1}.  
The corresponding value for the Sun during $1749$--$2017$ is $32\%$.  So, within our BL paradigm, 
we find that the observed tilt angle scatter is sufficient to account for the observed solar cycle variability.

\item[$\bullet$ ]
The simulation with the tilt saturation produces more variability than that with the flux saturation (compare Runs~C1 and B10).
Furthermore, the weaker diffusion in the CZ makes more variability (compare Runs~B5 and B10).

\item[$\bullet$ ]
The morphology of the magnetic fields in simulations with tilt scatters closely resembles observations.  
In particular, the surface radial field possesses more mixed polarity field (\Figs{fig:LDzoomed}{fig:hdflc}).

\item[$\bullet$ ]
With the inclusion of tilt scatter, the north-south asymmetry in the magnetic cycle is increased (\Fig{fig:zoomedBMR}).
However, the asymmetry never propagates for many cycles; through diffusion across the equator, 
the dynamo corrects this asymmetry within a few cycles. Similar behavior
is also observed in the Sun \citep[e.g.,][]{McInt13}.

\item[$\bullet$ ]
Tilt scatter also triggers grand minima and grand maxima.
The observed scatter of $\sigma=15^\circ$ for the recent cycles is
not sufficient to account for the grand minima inferred from
cosmogenic isotopes \citep{USK07}.  However, we do not include any 
positive feedbacks that
might enhance the scatter.  For example, weaker poloidal fields
will produce weaker toroidal fields that will in turn produce weaker
flux tubes with increased scatter due to buffeting by turbulent
convection.  

\item[$\bullet$ ] A larger scatter of
$\sigma=30^\circ$ leads to more frequent grand minima.  For example,
Run~B11 spends $18\%$ of its time in grand minima, compared to
$17\%$ for the Sun.  Larger scatter also increases the time spent
in grand maxima; $9.6\%$ for Run~B11 vs $9\%$ for the Sun.  

\item[$\bullet$ ]
Our model never shuts down at the observed tilt fluctuations, which was the case in the recent model of \citet{LC16}.

\item[$\bullet$ ]
The scatter in the tilt angle makes the dynamo slightly weaker in simulations
where the BMR delay distribution is fixed (compare Runs~A3--A4 and Runs~A5--A6).
However, this is not true in the cases of magnetic field-dependent delay distribution. 
The dynamo becomes even stronger with the increase of the tilt fluctuations; compare 
Runs B9--B11.

\item[$\bullet$ ]
In all simulations, we do not vary the meridional flow with time. 
Yet, we observe some variation in the cycle period. 
Particularly, the simulation with tilt fluctuations of $\sigma_\delta=15^\circ$ 
produces a variation
in the period which is indeed comparable to the observed solar cycle (\Fig{fig:pervsamp}).

\end{description}

As demonstrated in a 2D flux transport dynamo model by Karak \& Cameron (2016), we find that magnetic pumping enhances the efficiency of the dynamo.  In particular, the inclusion of magnetic pumping allows us to achieve sustained dynamo solutions using a BMR flux distribution comparable to the observed distribution ($\Phi_0 < 3$), even in the diffusion-dominated regime (see the B series in Table~1).  When magnetic pumping is not included, it is necessary to artificially boost the BMR flux ($\Phi_0 \gtrsim 28$) in order to achieve supercritical solutions (Runs~A3--A4).
Magnetic pumping also helps to make the magnetic field 
dipolar. The surface radial field, however, is largely dipolar (antisymmetric)
only during the solar minimum and 
it is dominated by 
quadrupolar (symmetric) mode during the solar maximum when several BMRs emerge at the surface to produce quadrupolar
 field (\Figs{fig:hdref}{fig:hdflc}(b)).
This type of multipolar surface magnetic field is in agreement with solar observations \citep{DBH12}.
Our dynamo model, however, can flip from the 
dipolar mode to the quadrupolar mode 
even with small parameter changes (e.g., changes in the BMR delay distribution; \Fig{fig:AB1}).

In our \bl\ model, as the poloidal flux produces the toroidal flux and then this toroidal flux
produces BMRs (\Fig{cartoon1}), the memory of the polar flux is largely reflected in the strength of the next sunspot cycle. We always obtain a strong correlation between the polar flux
and the sunspot of the next cycle (\Tab{table2}). However, the memory of the polar flux may not be propagated to multiple cycles. 
\citet{YNM08}, \citet{JCC07} and \citet{KN12} have shown that the memory of the polar flux is limited by the relative importance of diffusive and advective flux transport. However, here we show that it is also determined by the nonlinearity in the \bl\ process.
When the nonlinearity in BMR tilt is strong, the memory of the polar flux is limited to one cycle, irrespective of the flux transport.

\begin{acknowledgements}
We thank the anonymous referee for raising many constructive comments which helped to improve the presentation.
We are indebted to Lisa Upton for doing an internal review of this manuscript
and for providing the observed data of the BMR delay distribution used in \Fig{fig:tdelay}.
We extend our thanks to Arnab Rai Choudhuri for the discussion on the tilt angle fluctuations during Mark's visit to Bangalore.
We are also thankful to Gopal Hazra and Mausumi Dikpati for discussion.
BBK is supported by the NASA Living With a Star Jack Eddy Postdoctoral Fellowship Program,
administered by the University Corporation for Atmospheric Research.
The National Center for Atmospheric Research is sponsored by the National Science Foundation.
Computations were carried out with resources provided by NASA's High-End Computing program (Pleiades) and by NCAR (Yellowstone).
\end{acknowledgements}

\bibliographystyle{apj}
\bibliography{paper}

\end{document}

%% file: newcommands.tex
\newcommand{\EQ}{\begin{equation}}
\newcommand{\EN}{\end{equation}}
\newcommand{\EQA}{\begin{eqnarray}}
\newcommand{\ENA}{\end{eqnarray}}

\newcommand{\brac}[1]{\langle #1 \rangle}

\newcommand{\mps}{m~s$^{-1}$}
\newcommand{\kmps}{km~s$^{-1}$}

\newcommand{\cmss}{cm$^2$~s$^{-1}$}
\newcommand{\vpCZ}{$\gamma_{\rm CZ}$}
\newcommand{\vpS}{$\gamma_{\rm S}$}
\def\bl{BL}

\newcommand{\rhat}{\hat{\mbox{\boldmath $r$}} {}}
\newcommand{\thetahat}{\hat{\mbox{\boldmath $\theta$}} {}}
\newcommand{\phihat}{\hat{\mbox{\boldmath $\phi$}} {}}
\newcommand{\etas}{\eta_{\mathrm{S}}}
\newcommand{\etat}{\eta_{\mathrm{t}}}
\newcommand{\etab}{\eta_{\mathrm{CZ}}}
\newcommand{\etaRZ}{\eta_{\mathrm{RZ}}}

\newcommand{\Bsat}{B_{\rm sat}}

\newcommand{\taus}{\tau_{\rm s}}
\newcommand{\taup}{\tau_{\rm p}}

\newcommand{\Eq}[1]{Equation~(\ref{#1})}

\newcommand{\Sec}[1]{Section~\ref{#1}}
\newcommand{\Secs}[2]{Sections~\ref{#1} and \ref{#2}}
\newcommand{\Fig}[1]{Figure~\ref{#1}}
\newcommand{\Tab}[1]{Table~\ref{#1}}
\newcommand{\Figs}[2]{Figures~\ref{#1} and \ref{#2}}

%% file: paper.bbl
\begin{thebibliography}{85}
\expandafter\ifx\csname natexlab\endcsname\relax\def\natexlab#1{#1}\fi

\bibitem[{{Arlt} {et~al.}(2016){Arlt}, {Senthamizh Pavai}, {Schmiel}, \&
  {Spada}}]{Arlt16}
{Arlt}, R., {Senthamizh Pavai}, V., {Schmiel}, C., \& {Spada}, F. 2016, \aap,
  595, A104

\bibitem[{{Augustson} {et~al.}(2015){Augustson}, {Brun}, {Miesch}, \&
  {Toomre}}]{ABMT15}
{Augustson}, K., {Brun}, A.~S., {Miesch}, M., \& {Toomre}, J. 2015, \apj, 809,
  149

\bibitem[{{Babcock}(1961)}]{Ba61}
{Babcock}, H.~W. 1961, \apj, 133, 572

\bibitem[{Belucz {et~al.}(2015)Belucz, Dikpati, \& Forg\'acs-Dajka}]{beluc15}
Belucz, B., Dikpati, M., \& Forg\'acs-Dajka, E. 2015, 806, 169 (18pp)

\bibitem[{{Cameron} \& {Sch{\"u}ssler}(2015)}]{CS15}
{Cameron}, R., \& {Sch{\"u}ssler}, M. 2015, Science, 347, 1333

\bibitem[{{Cameron} {et~al.}(2013){Cameron}, {Dasi-Espuig}, {Jiang}, {I{\c
  s}{\i}k}, {Schmitt}, \& {Sch{\"u}ssler}}]{Ca13}
{Cameron}, R.~H., {Dasi-Espuig}, M., {Jiang}, J., {I{\c s}{\i}k}, E.,
  {Schmitt}, D., \& {Sch{\"u}ssler}, M. 2013, \aap, 557, A141

\bibitem[{{Cameron} \& {Sch{\"u}ssler}(2016)}]{CS16}
{Cameron}, R.~H., \& {Sch{\"u}ssler}, M. 2016, \aap, 591, A46

\bibitem[{{Charbonneau}(2010)}]{Cha10}
{Charbonneau}, P. 2010, Liv. Rev. Sol. Phys., 7, 3

\bibitem[{Charbonneau(2014)}]{C14}
Charbonneau, P. 2014, Ann. Rev. Astron. Astrophys., 52, 251

\bibitem[{{Charbonneau} \& {Dikpati}(2000)}]{CD00}
{Charbonneau}, P., \& {Dikpati}, M. 2000, \apj, 543, 1027

\bibitem[{{Chatterjee} {et~al.}(2004){Chatterjee}, {Nandy}, \&
  {Choudhuri}}]{CNC04}
{Chatterjee}, P., {Nandy}, D., \& {Choudhuri}, A.~R. 2004, \aap, 427, 1019

\bibitem[{{Choudhuri} {et~al.}(2007){Choudhuri}, {Chatterjee}, \&
  {Jiang}}]{CCJ07}
{Choudhuri}, A.~R., {Chatterjee}, P., \& {Jiang}, J. 2007, Physical Review
  Letters, 98, 131103

\bibitem[{{Choudhuri} \& {Karak}(2009)}]{CK09}
{Choudhuri}, A.~R., \& {Karak}, B.~B. 2009, Res. Astron. Astrophys., 9, 953

\bibitem[{{Dasi-Espuig} {et~al.}(2010){Dasi-Espuig}, {Solanki}, {Krivova},
  {Cameron}, \& {Pe{\~n}uela}}]{Das10}
{Dasi-Espuig}, M., {Solanki}, S.~K., {Krivova}, N.~A., {Cameron}, R., \&
  {Pe{\~n}uela}, T. 2010, \aap, 518, A7

\bibitem[{{DeRosa} {et~al.}(2012){DeRosa}, {Brun}, \& {Hoeksema}}]{DBH12}
{DeRosa}, M.~L., {Brun}, A.~S., \& {Hoeksema}, J.~T. 2012, \apj, 757, 96

\bibitem[{{Dikpati} \& {Charbonneau}(1999)}]{DC99}
{Dikpati}, M., \& {Charbonneau}, P. 1999, \apj, 518, 508

\bibitem[{{Dikpati} {et~al.}(2004){Dikpati}, {de Toma}, {Gilman}, {Arge}, \&
  {White}}]{Dik04}
{Dikpati}, M., {de Toma}, G., {Gilman}, P.~A., {Arge}, C.~N., \& {White}, O.~R.
  2004, \apj, 601, 1136

\bibitem[{{Dikpati} \& {Gilman}(2001)}]{DG01}
{Dikpati}, M., \& {Gilman}, P.~A. 2001, \apj, 559, 428

\bibitem[{{Dikpati} {et~al.}(2009){Dikpati}, {Gilman}, {Cally}, \&
  {Miesch}}]{Dik09}
{Dikpati}, M., {Gilman}, P.~A., {Cally}, P.~S., \& {Miesch}, M.~S. 2009, \apj,
  692, 1421

\bibitem[{{D'Silva} \& {Choudhuri}(1993)}]{DC93}
{D'Silva}, S., \& {Choudhuri}, A.~R. 1993, \aap, 272, 621

\bibitem[{{Fan} \& {Fang}(2014)}]{FF14}
{Fan}, Y., \& {Fang}, F. 2014, \apj, 789, 35

\bibitem[{{Fan} {et~al.}(1994){Fan}, {Fisher}, \& {McClymont}}]{FFM94}
{Fan}, Y., {Fisher}, G.~H., \& {McClymont}, A.~N. 1994, \apj, 436, 907

\bibitem[{{Featherstone} \& {Miesch}(2015)}]{FM15}
{Featherstone}, N.~A., \& {Miesch}, M.~S. 2015, \apj, 804, 67

\bibitem[{{Gilman} \& {Dikpati}(2000)}]{GD00}
{Gilman}, P.~A., \& {Dikpati}, M. 2000, \apj, 528, 552

\bibitem[{{Hale} {et~al.}(1919){Hale}, {Ellerman}, {Nicholson}, \&
  {Joy}}]{Hale19}
{Hale}, G.~E., {Ellerman}, F., {Nicholson}, S.~B., \& {Joy}, A.~H. 1919, \apj,
  49, 153

\bibitem[{{Hathaway} \& {Upton}(2016)}]{HU16}
{Hathaway}, D.~H., \& {Upton}, L.~A. 2016, Journal of Geophysical Research
  (Space Physics), 121, 10

\bibitem[{{Hazra} {et~al.}(2017){Hazra}, {Choudhuri}, \& {Miesch}}]{HCM17}
{Hazra}, G., {Choudhuri}, A.~R., \& {Miesch}, M.~S. 2017, \apj, 835, 39

\bibitem[{{Hazra} {et~al.}(2015){Hazra}, {Karak}, {Banerjee}, \&
  {Choudhuri}}]{Haz15}
{Hazra}, G., {Karak}, B.~B., {Banerjee}, D., \& {Choudhuri}, A.~R. 2015,
  \solphys, 290, 1851

\bibitem[{{Hazra} {et~al.}(2014){Hazra}, {Karak}, \& {Choudhuri}}]{HKC14}
{Hazra}, G., {Karak}, B.~B., \& {Choudhuri}, A.~R. 2014, \apj, 782, 93

\bibitem[{Hazra \& Nandy(2016)}]{hazra16}
Hazra, S., \& Nandy, D. 2016, 832, 9 (8pp)

\bibitem[{{Hotta} {et~al.}(2016){Hotta}, {Rempel}, \& {Yokoyama}}]{HRY16}
{Hotta}, H., {Rempel}, M., \& {Yokoyama}, T. 2016, Science, 351, 1427

\bibitem[{{Hotta} \& {Yokoyama}(2010)}]{HY10}
{Hotta}, H., \& {Yokoyama}, T. 2010, \apjl, 714, L308

\bibitem[{{Howard}(1991)}]{How91}
{Howard}, R.~F. 1991, \solphys, 136, 251

\bibitem[{Jackiewicz {et~al.}(2015)Jackiewicz, Serebryanskiy, \&
  Kholikov}]{jacki15}
Jackiewicz, J., Serebryanskiy, A., \& Kholikov, S. 2015, 805, 133 (9pp)

\bibitem[{{Jiang} {et~al.}(2014){Jiang}, {Cameron}, \& {Sch{\"u}ssler}}]{JCS14}
{Jiang}, J., {Cameron}, R.~H., \& {Sch{\"u}ssler}, M. 2014, \apj, 791, 5

\bibitem[{{Jiang} {et~al.}(2007){Jiang}, {Chatterjee}, \& {Choudhuri}}]{JCC07}
{Jiang}, J., {Chatterjee}, P., \& {Choudhuri}, A.~R. 2007, \mnras, 381, 1527

\bibitem[{Jouve \& Brun(2007)}]{jouve07}
Jouve, L., \& Brun, A. 2007, 474, 239

\bibitem[{{K{\"a}pyl{\"a}} {et~al.}(2016){K{\"a}pyl{\"a}}, {K{\"a}pyl{\"a}},
  {Olspert}, {Brandenburg}, {Warnecke}, {Karak}, \& {Pelt}}]{Kap16}
{K{\"a}pyl{\"a}}, M.~J., {K{\"a}pyl{\"a}}, P.~J., {Olspert}, N., {Brandenburg},
  A., {Warnecke}, J., {Karak}, B.~B., \& {Pelt}, J. 2016, \aap, 589, A56

\bibitem[{{Karak} \& {Brandenburg}(2016)}]{KB16}
{Karak}, B.~B., \& {Brandenburg}, A. 2016, \apj, 816, 28

\bibitem[{{Karak} \& {Cameron}(2016)}]{KC16}
{Karak}, B.~B., \& {Cameron}, R. 2016, \apj, 832, 94

\bibitem[{{Karak} \& {Choudhuri}(2011)}]{KC11}
{Karak}, B.~B., \& {Choudhuri}, A.~R. 2011, \mnras, 410, 1503

\bibitem[{{Karak} \& {Choudhuri}(2013)}]{KC13}
---. 2013, Res. Astron. Astrophys., 13, 1339

\bibitem[{{Karak} {et~al.}(2014{\natexlab{a}}){Karak}, {Jiang}, {Miesch},
  {Charbonneau}, \& {Choudhuri}}]{Kar14a}
{Karak}, B.~B., {Jiang}, J., {Miesch}, M.~S., {Charbonneau}, P., \&
  {Choudhuri}, A.~R. 2014{\natexlab{a}}, \ssr, 186, 561

\bibitem[{{Karak} {et~al.}(2015){Karak}, {K{\"a}pyl{\"a}}, {K{\"a}pyl{\"a}},
  {Brandenburg}, {Olspert}, \& {Pelt}}]{Kar15}
{Karak}, B.~B., {K{\"a}pyl{\"a}}, P.~J., {K{\"a}pyl{\"a}}, M.~J.,
  {Brandenburg}, A., {Olspert}, N., \& {Pelt}, J. 2015, \aap, 576, A26

\bibitem[{{Karak} \& {Nandy}(2012)}]{KN12}
{Karak}, B.~B., \& {Nandy}, D. 2012, \apjl, 761, L13

\bibitem[{{Karak} {et~al.}(2014{\natexlab{b}}){Karak}, {Rheinhardt},
  {Brandenburg}, {K{\"a}pyl{\"a}}, \& {K{\"a}pyl{\"a}}}]{Kar14b}
{Karak}, B.~B., {Rheinhardt}, M., {Brandenburg}, A., {K{\"a}pyl{\"a}}, P.~J.,
  \& {K{\"a}pyl{\"a}}, M.~J. 2014{\natexlab{b}}, \apj, 795, 16

\bibitem[{{Kitchatinov} \& {Olemskoy}(2011)}]{KO11}
{Kitchatinov}, L.~L., \& {Olemskoy}, S.~V. 2011, Astronomy Letters, 37, 656

\bibitem[{{Kitchatinov} {et~al.}(1994){Kitchatinov}, {Pipin}, \&
  {Ruediger}}]{KPR94}
{Kitchatinov}, L.~L., {Pipin}, V.~V., \& {Ruediger}, G. 1994, Astronomische
  Nachrichten, 315, 157

\bibitem[{{Komm} {et~al.}(1995){Komm}, {Howard}, \& {Harvey}}]{KHH95}
{Komm}, R.~W., {Howard}, R.~F., \& {Harvey}, J.~W. 1995, \solphys, 158, 213

\bibitem[{{Leighton}(1964)}]{Le64}
{Leighton}, R.~B. 1964, \apj, 140, 1547

\bibitem[{{Lemerle} \& {Charbonneau}(2017)}]{LC16}
{Lemerle}, A., \& {Charbonneau}, P. 2017, \apj, 834, 133

\bibitem[{{Lemerle} {et~al.}(2015){Lemerle}, {Charbonneau}, \&
  {Carignan-Dugas}}]{LCC15}
{Lemerle}, A., {Charbonneau}, P., \& {Carignan-Dugas}, A. 2015, \apj, 810, 78

\bibitem[{{Li}(2016)}]{Li16}
{Li}, D. 2016, ArXiv e-prints

\bibitem[{{Lopes} \& {Passos}(2009)}]{LP09}
{Lopes}, I., \& {Passos}, D. 2009, \solphys, 257, 1

\bibitem[{{McClintock} {et~al.}(2014){McClintock}, {Norton}, \& {Li}}]{MNL14}
{McClintock}, B.~H., {Norton}, A.~A., \& {Li}, J. 2014, \apj, 797, 130

\bibitem[{{McIntosh} {et~al.}(2013){McIntosh}, {Leamon}, {Gurman}, {Olive},
  {Cirtain}, {Hathaway}, {Burkepile}, {Miesch}, {Markel}, \&
  {Sitongia}}]{McInt13}
{McIntosh}, S.~W., {et~al.} 2013, \apj, 765, 146

\bibitem[{{McIntosh} {et~al.}(2015){McIntosh}, {Leamon}, {Krista}, {Title},
  {Hudson}, {Riley}, {Harder}, {Kopp}, {Snow}, {Woods}, {Kasper}, {Stevens}, \&
  {Ulrich}}]{Scott15}
---. 2015, Nature Communications, 6, 6491

\bibitem[{{Miesch} \& {Dikpati}(2014)}]{MD14}
{Miesch}, M.~S., \& {Dikpati}, M. 2014, \apjl, 785, L8

\bibitem[{{Miesch} {et~al.}(2012){Miesch}, {Featherstone}, {Rempel}, \&
  {Trampedach}}]{Miesch12}
{Miesch}, M.~S., {Featherstone}, N.~A., {Rempel}, M., \& {Trampedach}, R. 2012,
  \apj, 757, 128

\bibitem[{{Miesch} \& {Teweldebirhan}(2016)}]{MT16}
{Miesch}, M.~S., \& {Teweldebirhan}, K. 2016, \ssr

\bibitem[{{Mu{\~n}oz-Jaramillo} {et~al.}(2013){Mu{\~n}oz-Jaramillo},
  {Dasi-Espuig}, {Balmaceda}, \& {DeLuca}}]{Muno13}
{Mu{\~n}oz-Jaramillo}, A., {Dasi-Espuig}, M., {Balmaceda}, L.~A., \& {DeLuca},
  E.~E. 2013, \apjl, 767, L25

\bibitem[{{Mu{\~n}oz-Jaramillo} {et~al.}(2015){Mu{\~n}oz-Jaramillo},
  {Senkpeil}, {Windmueller}, {Amouzou}, {Longcope}, {Tlatov}, {Nagovitsyn},
  {Pevtsov}, {Chapman}, {Cookson}, {Yeates}, {Watson}, {Balmaceda}, {DeLuca},
  \& {Martens}}]{Mu15}
{Mu{\~n}oz-Jaramillo}, A., {et~al.} 2015, \apj, 800, 48

\bibitem[{{Nordlund} {et~al.}(2009){Nordlund}, {Stein}, \& {Asplund}}]{NSA09}
{Nordlund}, {\AA}., {Stein}, R.~F., \& {Asplund}, M. 2009, Living Reviews in
  Solar Physics, 6, 2

\bibitem[{{Olemskoy} \& {Kitchatinov}(2013)}]{OK13}
{Olemskoy}, S.~V., \& {Kitchatinov}, L.~L. 2013, \apj, 777, 71

\bibitem[{{Parfrey} \& {Menou}(2007)}]{PM07}
{Parfrey}, K.~P., \& {Menou}, K. 2007, \apjl, 667, L207

\bibitem[{{Parker}(1979)}]{Par79}
{Parker}, E.~N. 1979, {Cosmical magnetic fields: Their origin and their
  activity}

\bibitem[{{Passos} {et~al.}(2014){Passos}, {Nandy}, {Hazra}, \&
  {Lopes}}]{Pas14}
{Passos}, D., {Nandy}, D., {Hazra}, S., \& {Lopes}, I. 2014, \aap, 563, A18

\bibitem[{{Priyal} {et~al.}(2014){Priyal}, {Banerjee}, {Karak},
  {Mu{\~n}oz-Jaramillo}, {Ravindra}, {Choudhuri}, \& {Singh}}]{Priy14}
{Priyal}, M., {Banerjee}, D., {Karak}, B.~B., {Mu{\~n}oz-Jaramillo}, A.,
  {Ravindra}, B., {Choudhuri}, A.~R., \& {Singh}, J. 2014, \apjl, 793, L4

\bibitem[{{Rajaguru} \& {Antia}(2015)}]{RA15}
{Rajaguru}, S.~P., \& {Antia}, H.~M. 2015, \apj, 813, 114

\bibitem[{{Schatten} {et~al.}(1978){Schatten}, {Scherrer}, {Svalgaard}, \&
  {Wilcox}}]{Sch78}
{Schatten}, K.~H., {Scherrer}, P.~H., {Svalgaard}, L., \& {Wilcox}, J.~M. 1978,
  \grl, 5, 411

\bibitem[{{Schrijver} \& {Harvey}(1994)}]{SH94}
{Schrijver}, C.~J., \& {Harvey}, K.~L. 1994, \solphys, 150, 1

\bibitem[{{Senthamizh Pavai} {et~al.}(2015){Senthamizh Pavai}, {Arlt},
  {Dasi-Espuig}, {Krivova}, \& {Solanki}}]{pavai15}
{Senthamizh Pavai}, V., {Arlt}, R., {Dasi-Espuig}, M., {Krivova}, N.~A., \&
  {Solanki}, S.~K. 2015, \aap, 584, A73

\bibitem[{{Simard} {et~al.}(2016){Simard}, {Charbonneau}, \& {Dube}}]{SCD16}
{Simard}, C., {Charbonneau}, P., \& {Dube}, C. 2016, ArXiv e-prints

\bibitem[{{Solanki} {et~al.}(2008){Solanki}, {Wenzler}, \& {Schmitt}}]{SWS08}
{Solanki}, S.~K., {Wenzler}, T., \& {Schmitt}, D. 2008, \aap, 483, 623

\bibitem[{{Spruit}(1997)}]{Spr97}
{Spruit}, H. 1997, \memsai, 68, 397

\bibitem[{{Stenflo} \& {Kosovichev}(2012)}]{SK12}
{Stenflo}, J.~O., \& {Kosovichev}, A.~G. 2012, \apj, 745, 129

\bibitem[{{Upton} \& {Hathaway}(2014)}]{UH14}
{Upton}, L., \& {Hathaway}, D.~H. 2014, \apj, 792, 142

\bibitem[{{Usoskin}(2013)}]{Uso13}
{Usoskin}, I.~G. 2013, Living Reviews in Solar Physics, 10, 1

\bibitem[{{Usoskin} {et~al.}(2007){Usoskin}, {Solanki}, \& {Kovaltsov}}]{USK07}
{Usoskin}, I.~G., {Solanki}, S.~K., \& {Kovaltsov}, G.~A. 2007, \aap, 471, 301

\bibitem[{{Wang} {et~al.}(2015){Wang}, {Colaninno}, {Baranyi}, \&
  {Li}}]{Wang15}
{Wang}, Y.-M., {Colaninno}, R.~C., {Baranyi}, T., \& {Li}, J. 2015, \apj, 798,
  50

\bibitem[{{Wang} \& {Sheeley}(2009)}]{WS09}
{Wang}, Y.-M., \& {Sheeley}, N.~R. 2009, \apjl, 694, L11

\bibitem[{{Wang} \& {Sheeley}(1989)}]{WS89}
{Wang}, Y.-M., \& {Sheeley}, Jr., N.~R. 1989, \solphys, 124, 81

\bibitem[{{Yeates} {et~al.}(2008){Yeates}, {Nandy}, \& {Mackay}}]{YNM08}
{Yeates}, A.~R., {Nandy}, D., \& {Mackay}, D.~H. 2008, \apj, 673, 544

\bibitem[{{Zhang} {et~al.}(2010){Zhang}, {Wang}, \& {Liu}}]{ZWL10}
{Zhang}, J., {Wang}, Y., \& {Liu}, Y. 2010, \apj, 723, 1006

\bibitem[{{Zhao} \& {Chen}(2016)}]{ZC16}
{Zhao}, J., \& {Chen}, R. 2016, Asian Journal of Physics, 25

\end{thebibliography}
